\documentclass[manuscript]{aastex}
\usepackage{lscape}

\shorttitle{Spectroscopy of SMC Red Giants. I.}
\shortauthors{Parisi et al.}

\begin{document}

\title{\ion{Ca}{2} Triplet Spectroscopy of Small Magellanic Cloud Red Giants. I.
Abundances and Velocities for a Sample of Clusters }

\author{M.C. Parisi}
\affil{Observatorio Astron\'omico, Universidad Nacional de C\'ordoba}
\affil{Laprida 854, C\'ordoba, CP 5000, Argentina.}
\email{celeste@oac.uncor.edu}

\author{A.J. Grocholski}
\affil{Space Telescope Science Institute}
\affil{3700 San Martin Dr., Baltimore, MD 21218, USA.}
\email{aarong@stsci.edu}

\author{D. Geisler}
\affil{Departamento de Astronom{\'\i}a, Universidad de Concepci\'on}
\affil{Casilla 160-C, Concepci\'on, CP 4030000, Chile}
\email{dgeisler@astro-udec.cl}

\author{A. Sarajedini}
\affil{Department of Astronomy, University of Florida}
\affil{PO Box 112055, Gainesville, FL 32611, USA.}
\email{ata@astro.ufl.edu}

\and

\author{J.J. Clari\'a}
\affil{Observatorio Astron\'omico, Universidad Nacional de C\'ordoba}
\affil{Laprida 854, C\'ordoba, CP 5000, Argentina.}
\email{claria@oac.uncor.edu}

\begin{abstract}

We have obtained near-infrared spectra covering the \ion{Ca}{2} triplet lines
for a large number of stars associated with 16 SMC clusters using the VLT + FORS2.
These data compose the largest available sample of SMC clusters with
spectroscopically derived abundances and velocities.  Our clusters span
a wide range of ages  and provide good areal
coverage of the galaxy.  Cluster members are selected using a
combination of their positions relative to the cluster center as well as
their location in the CMD, abundances and radial velocities.  We determine mean cluster
velocities to typically 2.7 km s$^{-1}$ and metallicities to 0.05 dex (random
errors), from an average of 6.4 members per cluster.  By combining our
clusters with previously published results, we compile a sample of 25
clusters on a homogenous metallicity scale and with relatively small 
metalliciy errors, and thereby
investigate the metallicity distribution, metallicity gradient and
age-metallicity relation (AMR) of the SMC cluster system.  For all 25
clusters in our expanded sample, the mean metallicity [Fe/H] = $-$0.96 with $\sigma$
= 0.19. The metallicity distribution may possibly be bimodal, with
peaks at $\sim -$0.9 dex and $-$1.15 dex.  Similar to the LMC, the SMC
cluster system gives no indication of a radial metallicity gradient.
However, intermediate-age SMC clusters are both significantly more metal-poor
and have a larger metallicity spread than their LMC counterparts.
Our AMR shows evidence for 3 phases: a very early ($>11$ Gyr) phase in
which the metallicity reached $\sim -$1.2 dex, a long intermediate phase
from $\sim 10-3$ Gyr in which the metallicity only slightly increased,
and a final phase from 3$-$1 Gyr ago in which the rate of enrichment was
substantially faster.  We find good overall agreement with the model of
\citet{pag98}, which assumes a burst of star formation at 4 Gyr.
Finally,
 we find that the mean radial velocity of the cluster system is
148 km s$^{-1}$, with a velocity dispersion of 23.6 km s$^{-1}$ and no obvious signs of
rotation amongst the clusters.  Our result is similar to what has been
found from a wide variety of kinematic tracers in the SMC, and shows
that the SMC is best represented as a pressure supported system.
\end{abstract}


\section{Introduction}

The Small Magellanic Cloud (SMC) has long been recognized as being of
fundamental importance for a wide variety of astrophysical studies.
First, the current paradigm of galaxy formation suggests that spiral
galaxy spheroids, such as the Milky Way (MW) halo, are formed by the
accretion/merger of smaller, satellite galaxies (e.g., \citealt{sea78,zen03}).
 As one of the nearest low mass galaxies known,
the SMC is thus a possible prototype of pre-Galactic
fragments. 
However, current $\Lambda$CDM models suggest that the majority of the MW´s 
building blocks were assimilated very early in its history and that existing low
mass galaxies like the SMC are survivors of this process and thus underwent a
different chemical evolution (e.g., \citealt{rob05}).
At the least, many dynamical simulations
(e.g., \citealt{bek04}) suggest that the MW, Large Magellanic Cloud
(LMC) and SMC compose a longterm interacting system,
with the SMC and LMC likely to be eventually consumed
by the MW.  In addition, close encounters in the last few Gyr may well
have stimulated star formation on a global scale in
both the SMC and LMC (\citealt{mur80,gar94,yos03,bek04}). 
These early simulations predict bursts of star formation $\sim 0.2$Gyr ago 
that led to the formation of the eastern wing of the SMC and the Magellanic 
Bridge, and  a similar close encounter event some 4 Gyr ago.
Note however that the accurate proper
motions crucial for properly modeling the orbits of the SMC and LMC are
still problematic, and that 
the most recent values all suggest the Magellanic Cloud may be
unbound to each other and only beginning their first close encounter with the MW 
(\citealt{kal06,pia08}).  The SMC, with its low
global metallicity, is the best local counterpart to the host of distant dwarf
irregular and blue compact dwarf galaxies, making it an
attractive target for exploring the importance of metallicity in a
number of contexts, including star formation, initial mass function,
stellar and galaxy evolution, etc.

In particular, the star clusters of the SMC, because
they are (at least to first order)
simple stellar populations (SSPs), are an invaluable resource
with which we can explore the structure, kinematics,
star formation and chemical evolution history of the
SMC.  As a majority of stars
may have formed in clusters (see e.g., \citealt{lad03} for
the solar neighborhood and \citealt{cha06} for the SMC, but see \citealt{bas09} for an opposing view), their study
has become even more relevant.  On a cosmological
scale, star clusters in the SMC are of utmost importance for the study
and understanding of stellar populations in distant galaxies, since they
cover age and abundance space that is not occupied by their Galactic
counterparts. In addition, while the LMC cluster system
suffers from the well known age gap, where only one cluster is known to
have formed between $\sim$3 and $\sim$13 Gyr ago, (i.e.~during 3/4 of
its life; \citealt{dac91}, \citealt{gei97}), the SMC posesses clusters
that have been forming more or less 
continuously over the past $\sim$11 Gyr (e.g., \citealt{gl08b}).  The SMC
is the only dwarf galaxy in the Local Group that has formed and
preserved populous star clusters, without a significant age gap,
across (most of) the age of the Universe, and the production has been prolific; \citet{hod86}
estimates the SMC has some 2000 clusters.

Despite their utility and many clear advantages, SMC clusters have been surprisingly
underexploited. The number of clusters with well-determined ages from
deep main-sequence photometry is minimal.  Only two clusters have had their
detailed chemical abundances derived via high resolution spectroscopy
(NGC\,330 - \citealt{gon99}, \citealt{hil99}, NGC\,121 - \citealt{joh04}), 
and an additional six clusters have metallicities that
were derived from \ion{Ca}{2} triplet (CaT) spectroscopy of individual stars
(\citealt{dch98}, hereafter DH98).  Aside from these
eight clusters, all other existing abundance determinations are based
only on photometry or integrated spectroscopy.  Thus, the really detailed
information, viz. accurate ages and abundances, necessary to fully utilize
SMC clusters, both   as tracers of the SMC's formation and chemical evolution
history, and as templates for studying stellar populations in more
distant galaxies, is sorely lacking.

Our group has been working to ameliorate this situation over the past
several years.  Using Washington photometry, we have
derived ages and metallicities for almost 50 previously unstudied or poorly 
studied clusters \citep{pia01,pi05a,pi07a,pi07b,pi07c}.  These results
have greatly improved our understanding of the global properties of the
SMC cluster system,
provided constraints on the age-metallicity
relation (AMR), explored possible gradients, cluster formation
scenarios, etc.  However, these photometric cluster ages and
abundances suffer from two problems.  First, our photometry comes from
data obtained with a 1m class telescope, which is barely adequate to
reach the main sequence turn off for clusters older than several Gyr at
the distance of the SMC.  Second, while the photometry for the red giant
stars that are used to derive cluster abundances is generally good, the
Washington technique is known to require a significant age correction to
metallicities derived for intermediate-age ($\lesssim$ 5 Gyr) 
 objects \citep{gei03}, an age range that includes the majority of
the SMC clusters so far observed. This is of course illustrative of the infamous
age-metallicity degeneracy and is not a problem restricted to the
Washington system; while some age and metallicity estimates
from isochrone fitting to CMDs constructed in other
photometric systems exist, the degeneracy between age and metallicity
also makes these estimates inherently uncertain in the
absence of more solid metallicity measurements based on spectroscopic
data.

As mentioned above, spectroscopic based abundances only
exist for a handful of SMC clusters, with most of the sample coming from
the work by DH98.  They combined their CaT based metallicities with
ages from the literature to create the first accurate  AMR for the SMC, 
and found that it was consistent with a
simple closed box model and did not require the significant gas infall
or strong galactic winds that were needed to explain previous SMC AMR's
\citep[e.g.,][]{dop91}.  In addition, DH98 found that their cluster
velocities, like other kinematic tracers in the SMC, show no evidence
for any systematic rotation of the SMC.  However, their small sample
size and large age errors severely limit how well we can constrain the kinematics
and chemical evolution of the SMC.

While the CaT method does not measure Fe
abundances directly, previous authors have shown that the strength of
the CaT lines are an excellent proxy for [Fe/H], and can be used in
stellar populations covering a wide range of ages and abundances
\citep[e.g.,][]{col04,car07}. 
 An added benefit of using
the CaT is that it is very efficient; not only are RGB stars near their
brightest in the infrared, but the multiplexing capabilities of many
moderate-resolution spectrographs allows the observation of dozens of
stars simultaneously, greatly increasing the probability of identifying
cluster members.  In a previous paper \citep[hereafter G06]{gro06},
 we have applied this method to the LMC with excellent results,
following up on the pioneering  work by \citet{ols91}.
Using FORS2 on the VLT, we were able to identify more than 200 member
stars in 28 populous LMC clusters and determine accurate mean cluster
velocities ($\sigma$ = 1.6 km s$^{-1}$) and metallicities ($\sigma$ =
0.04 dex) and use these to explore the global cluster metallicity distribution, kinematics, etc.

To further refine our knowledge of the velocities and
abundances of clusters in the SMC, we here apply this powerful  technique
to the SMC, again using FORS2 on the VLT to
obtain medium-resolution near-infrared spectra of the CaT lines in 270
individual RGB stars in and around 16 SMC clusters.  Herein we present
our efforts to identify cluster members and derive mean abundances and
velocities for these  clusters.  In a future paper we will analyze the
several hundred field stars that were also observed as part of this
project.  This paper is organized as follows: In \S 2, we describe our
target selection process, and our spectroscopic observations and
reduction procedures are detailed in \S 3.  In \S 4 and \S 5, we present
the radial velocities and equivalent width measurements and the metallicity derivation,
respectively.  The membership selection process is described in \S 6 and
\S 7 compares our metallicity values with previous determinations. In \S 7 we
also discuss our metallicity results and in \S 8 the kinematics. Finally, in \S 9
we summarize our results.

\section{Target Selection}

To significantly improve upon the work of DH98, we have observed 16
SMC clusters.  Our cluster sample was chosen based on our Washington
photometry, where we have targeted clusters that appear to be at least
as old as 1 Gyr (this is approximately the minimum age for which CaT is
 well-calibrated), and cover as large a range in ages as possible
so as to sample the AMR
over a wide baseline.  We note that our clusters were chosen to be
complementary to those of DH98, as well as a similar study by Kayser et al. (2009, in preparation),
 whose targets we were aware of.
The ages of all but one of our clusters have been derived from Washington
photometry, using both the magnitude difference between the red clump (RC) and the
main sequence turn off \citep{gei97} as well
as fitting isochrones. For L\,17, we only had preliminary
Washington photometry and instead used the age derived by \citet{raf05}, 
which is based on integrated UBVI photometry.  We also attempted to select
 clusters
that were in relatively uncrowded fields and that were spread around the
galaxy so as to cover as wide an area and radial range as possible in order to
search for any global effects such as gradients.
Finally, we preferred clusters that were not too centrally condensed and
with a reasonable number of giants in order to allow us to observe at
least four definite cluster members from which to derive the cluster
metallicity.  In Table 1, we list our target clusters
along with their various catalog designations, right ascension,
and declination.

In Figure 1 we show the position of our cluster sample and the clusters
observed by DH98, in
relation to the optical center and bar, as well as several ellipses
aligned with the bar demonstrating our method for deriving the
galactocentric distance of a cluster (see section 7.3). The sample covers a significant
fraction of the galaxy.\\
DH98 studied 7 SMC 
clusters but they report the metallicity of only
6 of them, because NGC 361 did not have suitable photometry at the time
their work of DH98 was published.

$V$- and $I$-band pre-images of our target fields were taken by ESO
Paranal staff in 2005 August as part of Program 076.B-0533.
In all cases except one, the cluster was
centered on the upper (master) CCD, while the lower (secondary) CCD was
used to observe only field stars. In the case of BS\,121, the instrument
was rotated in order to center BS\,121 on the master CCD and include the nearby
cluster L\,72 on the secondary CCD.  Unfortunately, L\,72
turned out to be only a 25 Myr old cluster \citep{pi07a}, therefore,
all RGB stars observed around this cluster are actually SMC field
stars.  The pre-images were processed within IRAF \footnote{Image Reduction
and Analysis Facility, distributed by the National Optical Astronomy Observatories, which is
operated by the Association of Universities for Research in Astronomy, Inc., under contract
with the National Science Foundation.}
, and stars were
identified and photometered using the aperture photometry routines in
DAOPHOT \citep{ste87}.  Stars were cataloged using the FIND routine in
DAOPHOT and photometered with an aperture radius of 3 pixels. The $V$ and
$I$ band data were matched to form colors.  We note
that these exposures were purposefully lengthened in order to go
deep enough to reach below the main sequence turnoff
for most of our clusters and in a future paper we will
present the full point spread function fitting photometry and the
resulting cluster ages, based on isochrone fitting to our data.

Spectrocopic targets were chosen based on the instrumental CMD, giving
highest priority to stars lying along the apparent cluster giant branch. Each
candidate was visually inspected to ensure location within the cluster
radius (judged by eye) and freedom from contamination by very nearby
bright neighbors. In each cluster we then looked for maximum packing of
the $\approx 8"$-long slits into the cluster area and for the best
possible coverage of the magnitude range from the horizontal branch/red
clump ($V \approx$ 19.5) to the tip of the RGB ($V \approx$ 17) in order
to select the best cluster targets. The positions of each target were
defined on the astrometric system of the FORS2 pre-images so that the
slits could be centered as accurately as possible, and the slit
identifications were defined using the FORS Instrument Mask Simulator
(FIMS) software provided by ESO; the slit masks were cut on Paranal by
the FORS2 team. In general we were able to place slits on some 10 stars
per cluster that we judged initially to be likely members based on their location
and photometry. This left ample additional
space for other targets, both on the master chip and especially on the
secondary chip, and we selected as many field giants as possible in
order to explore their metallicities and kinematics as well. Typically we
observed about 25 probable field giants around each cluster.

\section{Spectroscopic Observations and Reductions}

The spectra of the program stars were obtained during 2005 November in
service mode by the VLT staff, using the FORS2 spectrograph in mask
exchange unit (MXU) mode.  Our instrumental setup was identical to that
discussed in G06, which can be consulted for a more detailed
description.  We used slits that were 1'' wide and 8'' long.
 In all fields, we obtained a single 900 s exposure with
a typical seeing of less than 1''. Pixels were binned 2x2, yielding a plate scale
of 0.25'' pixel$^{-1}$. The spectra have
a dispersion of $\sim$ 0.85 \AA/pixel (which corresponds to a resolution
of 2-3 \AA)  with a characteristic rms scatter of $\sim$ 0.06 \AA,
and cover a range of $\sim$ 1600 \AA \space in the region of
the CaT (8498 \AA, 8542 \AA \space and 8662 \AA). S/N values ranged from
 $\sim$ 10 to $\sim$ 70  pixel$^{-1}$. Calibration exposures, bias frames
and flat-fields were also taken by the VLT staff.\\

We followed the
image processing detailed in G06. In brief, we used the IRAF 
 task {\it
ccdproc} to fit and substract the overscan region, trim the images, fix
bad pixels, and flat-field each image. We then corrected the images for
distortions, which rectifies the image of each slitlet to a constant
range in the spatial direction and then traces the sky lines along each
slitlet and puts them perpendicular to the dispersion direction.  We
used the task {\it apall} to define the sky background and extract the
stellar spectra onto one dimension. The tasks {\it identify, refspectra}
and {\it dispcor} were used to calculate and apply the dispersion
solution for each spectrum. Finally, the spectra were
continuum-normalized by fitting a polynomial to the stellar continuum.
Two examples of the final spectra in the CaT region can be seen in
Figure 2. Clearly, the three Ca II lines dominate the red giant's
spectra in this region. Also note the stronger Ca lines in the bottom spectrum, 
which, given the very similar $T_{eff}$ and log g values (as indicated by their
virtually identical $v-v_{HB}$ values), graphically illustrates the higher 
metallicity of the bottom spectrum and the power of the CaT technique.

\section{Radial Velocity and Equivalent Width Measurements}

Radial velocities (RVs) of our target stars are useful for evaluating cluster membership
since it is expected that SMC stars have substantially higher RVs than 
foreground MW stars and that
the cluster's velocity dispersion should be
small in comparison to that of the surrounding SMC field
stars. In addition, the program used to measure the EWs of the CaT
lines needs to know the RV in order to make the Doppler correction and
to derive the CaT line centers.

To measure RVs of our program stars, we performed
cross-correlations between their spectra and the spectra of 32 bright
Milky Way open and globular cluster template giants using the IRAF task
{\it fxcor} \citep{ton79}. We used the template stars of \citet[hereafter C04]{col04}
 who observed these stars with a
setup very similar to ours. The template spectra correspond
to 2 stars each from Melotte\,66 and NGC\,2298, 3 stars each from NGC\,4590, 
Berkeley\,20 and 47\,Tuc, 6 stars each from NGC\,1904 and Melotte\,67
and 7 stars from Berkeley\,39. 
In addition, {\it fxcor}
makes the necessary correction to place the derived RV in the
heliocentric reference frame. We adopted the average of each
cross-correlation result as the heliocentric radial velocity of a star,
finding a typical standard deviation of $\sim$ 6 km s$^{-1}$.

As shown in previous papers \citep[e.g.,][]{irw02}, errors in
centering the image in the spectrograph slit can result in inaccuracies
in determining RVs.  In order to correct for this, we measured the
offset $\Delta x$ between each star's centroid and the
corresponding slit center by inspecting the through-slit image taken
immediately before the spectroscopic observation, according to the
procedures described by C04 and G06.  Then, the velocity correction
$\Delta v$ will be:

\begin{equation}
\Delta v = \Delta x \frac{dv}{dx} ,
\end{equation}

\noindent where $\Delta x = x_{cen}(slit) - x_{cen}(star)$ and $dv/dx$ is our
spectral resolution.

 Typical centering errors are no larger than 0.90 pixels 
and our measurement precision is estimated to be ~0.15 pixels. 
Given our spectral resolution of $\sim$ 30 km
s$^{-1}$ px$^{-1}$, the velocity corrections we have
applied range from $\vert\Delta v\vert$ = 0 to 27 km s$^{-1}$ and the
typical error introduced in the RV turns out to be $\pm$ 4.5 km s$^{-1}$. This
error, added in quadrature to the one resulting from the
cross-correlation, yields a total error of 7.5 km s$^{-1}$. This value
has been adopted as the typical RV error (random $+$ systematic)
of an individual star.

To measure EWs we have used a previously written
FORTRAN program (see C04 for details).  The region around the CaT is
highly contaminated by weak absorption lines, so we follow the procedure
of \citet{arz88} and define continuum bandpasses on both
sides of each CaT line.  The rest wavelength of the CaT line centers,
along with the continuum bandpasses that we have adopted from \citet{arz88}, 
are listed in Table 2 of G06.  For RGB stars, the
flux in the wavelength range around the CaT is virtually flat, so the
``pseudo-continuum'' for each CaT line is then easily determined by a
linear fit to the mean value in each pair of continuum windows. The
``pseudo-equivalent width'' is calculated by fitting a function to each
CaT line with respect to the pseudo-continuum.  Following C04 and G06,
we define our metallicity index, $\Sigma W$, as the sum of the EWs of
the three CaT lines.  As discussed by \citet{ru97b} and C04,
just fitting a Gaussian (G) function to the very strong CaT lines
tends to underestimate the strength of the wings of the
line profile, especially at high metallicity. C04 found that, by adding a
Lorentzian profile to the Gaussian (GL), it is possible to maintain
sensitivity over the full range of abundances.  However, C04 also showed
that, by adding artificial noise to one of their spectra, for stars with
S/N less than $\sim$ 15-20, there is no significant difference in the EWs
measured using the two profile shapes.  In light of this, for spectra with
S/N $>$ 20, we use the composite line fits (GL), but for spectra with S/N $<$
20, we prefer a Gaussian-only fit, since it has fewer free parameters.
We correct the Gaussian-only fit for the low
S/N spectra as follows.  We measured $\Sigma W$ for the spectra with
high S/N ($>$ 20) using both a Gaussian ($\Sigma W_G$) and a
Gaussian+Lorentzian ($\Sigma W_{GL}$) function.  In Figure~3 we plot $\Sigma W_G$
 vs.~$\Sigma W_{GL}$, and find a best-fit linear relation:

\begin{equation}
\Sigma W_{GL} = \frac{\Sigma W_G - 0.280}{0.848} ,
\end{equation}

\noindent with a scatter of 0.23 \AA. Then, for low S/N ($<$ 20)
spectra, we measured the EW using G-fits, calculated $\Sigma W_G$ and
then calculated via equation (2) the corresponding $\Sigma W_{GL}$.
Errors in the EW measurements were estimated by measuring the rms
scatter of the data about the fits, and yielded typical errors of $\sim$
0.1-0.5 \AA \space depending on the line and the S/N of the spectra. To
calculate the errors involved in the $\Sigma W$ calculation, we added in
quadrature the error of the EW of each of the three lines. In the case of those
spectra with S/N $<$ 20 this last error was added in quadrature with the
scatter of equation (2). Note that C04 also derived a linear relation between
$\Sigma W_G$ and $\Sigma W_{GL}$ using galactic open and globular
cluster stars. They found a change in the slope at $\Sigma W_{GL}$
$\approx$ 6.5 \AA.  There is no strong evidence of this slope change in
our data.

\section{Metallicities}

Several studies have calibrated the relationship between the strengths
of the CaT lines and stellar abundance. In all cases, the selected CaT
index uses a linear combination of the EW of two or three individual Ca
II lines to form the line strength index $\Sigma W$.
For example, \citet{sun93} and \citet{col00} used a linear combination 
of the two strongest lines
(8542 \AA ~and 8662 \AA), while in other cases all
three lines were used, but each line was assigned a different weight in
deriving the sum \citep[e.g.,][]{ru97a}.  Since
our spectra are of high enough quality that all three CaT lines are well
measured, we adopted for $\Sigma W$ the same definition as C04, in which
all three lines are used with equal weight, namely:

\begin{equation}
\Sigma W = EW_{8498} + EW_{8542} + EW_{8662},
\end{equation}

Although theoretical and empirical studies have shown
that effective temperature, surface gravity, and [Fe/H] all play a role
in CaT line strengths \citep[e.g.,][]{jor92,cen02}, \citet{ard91} showed that there
is a linear relationship between a star's absolute magnitude and
 $\Sigma W$ for red giants of a given metallicity.  Following  previous
authors, we then define a reduced equivalent width, $W'$, to remove the effects
of surface gravity and temperature on $\Sigma W$ via its luminosity 
dependence:

\begin{equation}
W' = \Sigma W + \beta (V-V_{HB}), 
\end{equation}

\noindent in which the introduction of the difference between the visual
magnitude of the star ($V$) and the cluster's
horizontal branch/red clump ($V_{HB}$) also removes any
dependence on cluster distance and interstellar reddening. The value of
$\beta$ has been investigated by previous authors and
we have adopted the value obtained by C04: $\beta$ = 0.73 $\pm$ 0.04.
We chose C04's value of $\beta$ because they used both globular and open
clusters, covering a range of ages similar to the clusters in our sample,
to derive it and their instrumental setup is very similar to
ours.  We note that, as discussed in detail in G06, the
use of C04's value for $\beta$, combined with defining the brightness of
our target RGB stars relative to the HB of their parent cluster, incorporate
any age effects into the CaT calibration, thus it is not necessary to
make any corrections for the ages of our clusters.  As shown by
\citet{ru97a}, there is a linear relationship between a
cluster's reduced EW and its metallicity on the \citet{cag97} abundance scale 
for globular clusters of the Milky Way. C04
extended this calibration to a wider range of ages (2.5 Gyr $\leq$ age
$\leq$ 13 Gyr) and metallicities ($-2.0 \leq$ [Fe/H] $\leq -0.2$)
by combining the metallicity scales of \citet{cag97} and \citet{fri02} for globular and open clusters,
respectively. Although some of our clusters are
younger than 2.5 Gyr, we adopted the C04 relationship,

\begin{equation}
[Fe/H] = (-2.966 \pm 0.032) + (0.362 \pm 0.014)W',
\end{equation}

\noindent to derive the metallicities of our entire cluster
sample.  \citet{car07} investigated the behaviour of CaT in the
age range
of 0.25 Gyr $\leq$ age $\leq$ 13 Gyr and the metallicity range $-2.2
\leq$ [Fe/H] $\leq +0.47$.  In
spite of the extended age range of their calibration, we decided to use C04's calibration
for three reasons.  First, the calibration of \citet{car07} uses the 
absolute magnitude of a target star
(rather than the brightness relative to the HB, as we have adopted), thus
requiring accurate photometric calibration and
distance to the star.  Unlike the
LMC, the SMC and its cluster system is substantially extended along the line of
sight
\citep[e.g.,][]{cro01,gl08b}, making it difficult to determine the
distance to an individual cluster, thereby introducing an additional
source of error in our calculation.  Second, using $(V-V_{HB})$ instead
of absolute magnitude removes the need to make any assumptions about the
foreground reddening toward a cluster.  Finally, \citet{car07} performed 
estimates of the differences in $W'$ as a function of age
using the \citet{jor92} models and {\it BaSTI} stellar
evolution models \citep{pie04}. From their data, they
confirmed that the influence of age is small, even for ages $<1$ Gyr.
Based on this result, we are confident that the C04
calibration can be accurately extended to younger ages.  Note that our
youngest clusters are estimated to be $\sim$ 0.9 Gyr (Lindsay\,106 and
108), but the rest are \raise-0.5ex\hbox{$\buildrel>\over\sim$} 1 Gyr.

In addition, \citet{bat07} have shown that [Fe/H] values
derived from CaT agree with those from high resolution spectroscopy and
detailed model atmosphere analyses to within 0.1 - 0.2 dex over the
range $-2.5\leq$ [Fe/H]$\le -0.5$, based on FLAMES data for a large sample
of stars in the Sculptor and Fornax dwarf spheroidal galaxies. It is
well known that such stars have distinct [Ca/Fe] ratios from those in
the Galactic globular cluster calibrators \citep[e.g.,][]{she01,gei07}
 and also possess a larger range of stellar ages, extending
to much smaller
values. Nevertheless, CaT yields metallicities that are very close to
real Fe abundances, reinforcing our confidence in the technique.
Our estimate of the total
 metallicity error (random $+$ systematic) per star ranges from 0.09 to 0.35 dex, with a
mean of 0.17 dex.

\section{Membership Selection}

We generally follow the procedure described in detail
by G06 to determine cluster membership, and give a brief description
herein.  As it is necessary to measure the HB/RC in each cluster, we
begin by determining each cluster's radius which will allow us to
separate the cluster stars from the field to first order.  One of the most robust ways
to evaluate the cluster radius is using the stellar density radial
profile, which we constructed following the method described by \citet{pi07c}.
We start by determining the center of the cluster by
building projected histograms in the x and y directions using the coordinates
from the aperture photometry.  Once this is
performed, we fit Gaussian functions to the distributions using the task
{\it NGAUSSFIT} of {\it IRAF}, adopting the derived values as the
coordinates of the cluster's center. Finally we constructed the cluster
radial profile based on star counts within boxes of 50 pixels on a side
distributed throughout the whole field of each cluster. The number of
stars per unit area at a given radius can be calculated through the
expression given by \citet{pi07c}:

\begin{equation}
\frac{n_{r+25} - n_{r-25}}{(m_{r+25} - m_{r-25})50^2},
\end{equation}

\noindent where n$_r$ and m$_r$ are the number of stars and boxes in a circle of
radius r, respectively.  We adopted as the cluster radius the value where the gradient in the density profile significantly flattened. In Figure~4, we use L\,19 to
illustrate the process employed for all clusters. The x-axis represents
the radial distance and the y-axis the number of stars (N) per unit
area. The vertical line on the profile represents the
adopted cluster radius, which is used in the
subsequent analysis.  Note that the
adopted cluster radius can differ from the
more typical definition, in which the radius is
the distance from the center to the point
where the stellar density profile intersects the background \citep{pi07c}. 
As is seen in Figure 4, the background
level is $\sim$ 0.025 stars/arcsec$^2$. Thus, the real cluster radius
probably extends farther than the adopted one.  We decided to use for
our analysis a conservative radius in order to maximize the probability
of cluster membership.

After defining the cluster radii, we build CMDs for
each cluster using aperture photometry that is derived from our $v$- and
$i$-band preimages.  From here on, we denote our photometry with lower
case letters to emphasize that these are presently uncalibrated instrumental
values.  We are able to make use of the instrumental photometry due to
the fact that the $V-I$ color term is expected to be small in the FORS2
filter system ($<$ 0.02 mag), and should therefore have little effect
over the small range in colors covered by the RGB stars.  Due to the
fact that all of our clusters are too young and metal rich to have a fully
formed HB, we instead adopt the median value of the core helium-burning red
clump stars. The estimated change in $v_{HB}$  from
the old to the young populations is on the order of 0.05 dex \citep{gro06} which is
smaller than our measurement errors. All of the
clusters in our sample show RC features that are well populated.  We
measure $v_{HB}$ as the median value of all stars inside of a box that
is 0.7 mag in $v$ and 0.3 mag in $v-i$ and centered on the RC by eye.
We use the median value instead of the mean since it reduces the effect
of outliers on our calculations.  Errors in $v_{HB}$ are taken as the
standard error of the median.

In Figure~5 we plot the $xy$ positions of all stars
photometered in and around L\,19, and the corresponding CMD is plotted in
Figure~6.  Our spectroscopic targets are represented by the large filled
symbols, and the adopted cluster radius is marked by the large circle in
Figure~5.  Target stars marked in blue are considered non-members due to
their location outside of the adopted cluster radius but were still observed 
in order to study the SMC field population.

To further discriminate member from non-member stars,
we can take advantage of the behavior of the radial velocities and
metallicities as functions of clustercentric distance.  Cluster members
should have a smaller velocity dispersion than field stars and
may also have a mean velocity different from
the field.  Figure 7 shows how RV varies as a function
of distance for L\,19.  We have adopted an intrinsic cluster
velocity dispersion of 5 km s$^{-1}$, appropriate for typical Magellanic Cloud clusters \citep{pry93}, 
which added in quadrature with our
adopted radial velocity error of 7.5 km s$^{-1}$, yields an expected
dispersion of $\sim$ 9 km s$^{-1}$. We have rounded this up and adopted
$\pm$ 10 km s$^{-1}$ for our RV error cut, represented by horizontal
lines in Figure 7. The vertical line represents the adopted cluster
radius. Figure 8 plots metallicity versus distance from the cluster
center for L\,19.  We have adopted an  [Fe/H] error cut of
$\pm$0.20 dex (horizontal lines) given our estimate of a typical
metallicity error of 0.17 dex
for an individual star.
Again, the vertical line in Figure 8
marks the cluster radius. In both plots blue symbols represent
non-members that are outside the cluster radius, teal and green symbols
represent non-members that we eliminated because of discrepant radial
velocity and metallicity, respectively, and red symbols
are targets that have passed all three cuts and are
therefore considered cluster members. This procedure is followed for
each cluster in our sample.

In Table 2 we list the information for the member stars. Columns (1), (2)
and (3) show the identification of the star,
right ascension and declination respectively.  Table 2 also lists
heliocentric radial velocity and its error in columns (4) and (5),
$v-v_{HB}$ in column (6), $\Sigma W$ and its error in columns (7) and
(8) and metallicity and its error in column (9) and (10).  
In Figure 9 we give the $\Sigma W$ vs. $v-v_{HB}$ plot for all cluster's
members. Dashed lines
represent isometallicity lines of [Fe/H] = 0, $-$0.5, $-$1, $-$1.5 and $-$2 (from top to bottom).

Finally, using these member stars, we calculated the mean cluster
metallicity and its standard error of the mean. The final results are
given in Table 3 where we list successively: cluster name, the number of
stars, n, established as members, mean heliocentric
radial velocity with its error, the mean metallicity followed by its
error, position angle and the
semi-major axis $a$ (see section 7.3 for details about the calculation of $a$).
Both radial velocity and metallicity errors correspond to the
standard error of the mean (s.e.m.). We derive the mean cluster metallicity (mean s.e.m., random error) to
0.05 dex and the mean radial velocities to 2.7 km s$^{-1}$ for an average of 6.4 members
per cluster. This compares well to the mean values we derived for our
LMC cluster sample: 0.04 dex and 1.6 km s$^{-1}$ for 8 stars, and is significantly
better than the mean values obtained by DH98: 0.12 dex and 4.7 km s$^{-1}$ for 4
stars per cluster (in a total of 6 clusters).\\

It is important to emphasize   that the metallicity determination of BS\,121 is
more uncertain than for the other clusters
due to the fact that it was difficult to select convincing metallicity cuts. The stars selected via
radial velocity   appeared to belong to two groups.
The stars of the first group have metallicities
similar to the field metallicity in that region ($\sim -$0.95 dex) and
are probably field stars.   We therefore assumed that the lower metallicity group are the actual cluster members and used them 
to calculate the metallicity of BS\,121. It is  necessary to observe
more stars in order to improve our                  metallicity uncertainty.

\section{Metallicity Results}

\subsection{Comparison to Previous Metallicity Determinations and Sample
Enlargement}

Before investigating the implications of our metallicity results, it is first important
to see how they compare with any previous determinations.
Since our cluster sample is limited to only 15 clusters, in order to study
global effects, obtain better statistical errors, etc., it would also be beneficial
to enlarge the sample size as much as possible by including other studies
whose metallicities we find to be on a similar scale to ours and also have 
relatively small errors.

No previous CaT or high resolution spectra have been obtained for
any of our clusters. The DH98 study, of course, includes several other
clusters studied via
CaT so we will add     their sample to ours.
\citet{pi05b} did derive
reddening and ages of L\,5 and L\,7 using low-resolution integrated spectra.
However,                    the metallicities       they report are
not derived directly from their spectra.

Metallicities for a number
of our clusters have been determined photometrically.
\citet{pi05a} presented results on
the age and metallicity estimates of most clusters of our sample (BS\,121,
HW\,47, HW\,84, HW\,86, L\,4, L\,5, L\,6, L\,7, L\,19 and L\,27) calculated from
CMDs in the C and T$_1$ bands of the
Washington system.  The same technique was used by \citet{pi07b}
to derive the corresponding parameters for L\,110.

Table 4 summarizes the previous photometric metallicity and age
determinations for our clusters. Column (1) shows the cluster name, the
metallicity and its respective error are    found in column (2) and the
age in column (3). Column (4) lists the reference from where the
metallicities and ages were taken. There is no previous metallicity
determination for the clusters L\,17, L\,72, L\,106 and L\,108.  Note that we have
used the ages derived from Washington photometry in all cases except
L\,17. These ages are based on main sequence photometry and have been
determined using the same technique, thus assuring that they are on a
homogeneous scale.

In Figure 10 we have plotted the metallicities derived by \citet{pi05a,pi07b}
 as a function of the metallicities from the present work
for the 11 clusters in common. We note that their metallicities for these clusters were
derived by comparing the cluster red giant branches with those of fiducial
globular clusters from \citet{gei99} and then correcting the photometric metallicities for their age dependence via the prescription given in \citet{gei03}.
The mean difference between our metallicities and those derived by
\citet{pi05a,pi07b} is $-0.04 \pm 0.25$ dex, with our values
being more metal-rich, while a mean absolute difference of 0.20 $\pm$
0.14 dex, for the 11 clusters in common, is found. The mean errors of the spectroscopic and photometric
metallicities are 0.05 and 0.26 dex, respectively. Then the mean
expected error (sum in quadrature of these two errors) is 0.27 dex,
which is larger than the observed difference. Thus, our results are in good
agreement with those derived from the Washington system. This indicates
that the Washington photometric technique employed by \citet{pi05a,pi07b}
 is well calibrated and especially that their correction
for age effects is also appropriate. We further tested    this latter point
by plotting the difference in metallicity between the two studies vs. age (Figure 11).
There is no systematic
trend evident and the differences are within the range expected from the
given errors except for BS\,121, whose CaT metallicity is problematic.
We conclude that the age correction procedure used in the Washington technique,
outlined in \citet{gei03}, is indeed appropriate.

Although our spectroscopic metallicities and the photometric
metallicities of \citet{pi05a,pi07b} are consistent with each other, it
is important to note that our CaT-based abundances are much more precise.
As shown by the error bars in Fig. 10, our errors are $<$ 0.07 dex while
the photometric errors range from 0.2 to 0.4 dex.  While this comparison
gives us confidence that the photometric abundances from the Washington
technique are on a similar scale to ours, the associated errors mean that
their data are not precise enough to serve our purposes.

However, we can enlarge our sample size by
including metallicities derived in several other studies which we deem to be
essentially on the
same scale as ours, 
although we have essentially no clusters in common.
These studies include DH98 and any other clusters studied with the CaT 
technique, as well as any clusters studied with high resolution spectroscopy. 
For the DH98 clusters, we use the values
they give in their Table 3 on the Caretta \& Gratton metallicity scale and uncorrected
for age effects, as these are the closest to our scale.
Note that their technique is not identical to ours as they used the
sum of only the 2 strongest Ca lines, but the metallicities derived
should be very similar.
\citet{gl08b} also give CaT metallicities for 3 additional clusters. We
use their values on the \citet{cag97} scale to match ours. Note
that the errors associated with the mean cluster metallicities derived by both 
DH98 and Glatt et al. range from 0.06 -- 0.13 dex, with a mean of 0.10 dex, 
twice as large as our mean error. However, we feel that these are good enough
for our purposes. Finally, the only cluster with a metallicity based on high
resolution spectroscopy published in a refereed journal to date is NGC 330
\citep{gon99}, who derive $-0.94 \pm 0.02$ for this very young
cluster.  We also add this derivation
to our sample.
This gives a total sample of 25 clusters with metallicities on a reasonably homogeneous
scale and with relatively small errors.

\subsection{Metallicity Distribution}

The cluster metallicity distribution (hereafter MD) is an important diagnostic of the global
chemical evolution of a galaxy and is useful for an overall comparison of the
cluster systems of different galaxies.
First, we derive a mean metallicity for our CaT sample of $-$0.94, with $\sigma
$ (standard deviation) $= 0.19$, while for the full sample these values become $-$0.96 and 0.19. 
The mean values are in very good agreement with each other and the global mean
value of $-1$ found by \citet{car08} from CaT spectra of a large
number of field giants.

However, we note here that the ages for these clusters are not all on
the same scale. Those for our sample are derived
from Washington main sequence photometry but in some cases are based on isochrones and
others based on the magnitude difference between the RC and turnoff, although
these have been shown to give similar ages for clusters in common \citep{gei03}. 
DH98 adopted the ages given in the original CMD references for their cluster
sample, which are based on MS photometry but vary in the details such as
photometric system, age calibration, etc. \citet{gl08b} use deep HST ACS
photometry. Thus, the ages of our full sample
are on a more heterogeneous scale than the metallicities. We are in the process of
obtaining better ages for our own clusters based on a more sophisticated data
reduction and analysis, i.e. psf-photometry of the (long exposure) preimages, but this is beyond
the scope of the present paper. 
For this reason, we restrict our investigation
utilizing the ages of our clusters to only the AMR. We feel it is premature to use
the current age estimates to investigate properties that demand  more
homogeneous data, e.g., the fascinating question of whether or not there have
been bursts of cluster formation, as first suggested by \citet{ric00} and more
recently by \citet{pi05a}. Such analysis requires a larger and more
homogeneous sample than currently available.

We can divide our sample into two age bins: younger and older than 3 Gyr. The
12 older clusters have a mean age of 5.8 Gyr and a mean
metallicity of $-1.08, \sigma = 0.17$ (0.049 s.e.m.),
 while the 13 younger clusters have values of 1.6 Gyr, $-0.85, 0.15$ (0.041 s.e.m.). The two mean metallicities
differ by 0.23 $\pm$ 0.06 dex (3.6 sigma). We feel, despite the caveats given above, that this
difference is probably real and simply reflects the AMR discussed in detail
below, in which we find that younger clusters are generally slightly more
enriched.
Note that the mean metallicity of even the younger division is still somewhat lower
than that generally attributed to the present-day galaxy of $\sim -0.65$
\citep[e.g.,][]{rus89,hil97} as it includes objects of a few Gyr 
in age and there does appear to have
been significant enrichment after  this time \citep[e.g.,][and below]{har04,car08}.
Particularly interesting in this regard is the youngest cluster in our combined
sample, NGC 330, which at an age of $30 Myr$ has a metallicity of --0.94.

The MD is shown in Figure 12 (top panel), in which we also show the MD
 for LMC clusters derived by G06 in our CaT study (bottom panel). The SMC clusters fall             in a
rather small metallicity range of $<0.8$ dex, from $-$0.6 to $-$1.4, and  are concentrated
in the 0.5 dex range from $-$0.75 to $-$1.25. This is unlike the case of LMC
clusters, which cover $\sim 2$ dex in metallicity (G06), with higher and lower metallicities than found
in their SMC counterparts. G06 found that the older LMC clusters in their
sample of 28 (very comparable to the present sample size) had metallicities 
(also based on CaT) from
$\sim -1.3$ to $-$2.1 but that the younger clusters, which formed after the age gap
ended some 3 Gyr ago, have a very tight MD, with a mean [Fe/H] $=-0.48,
\sigma = 0.09$. The broad characteristics of the MD
 of the cluster systems of the Magellanic Clouds  have been known for
some time. \citet{dac91}  noted that (sic) ``The LMC managed to make metals
but no clusters during the age gap while the SMC managed to make clusters but no
metals". These curious facts are now even more evident and corroborated in
much greater detail than known at that time, but their explanation remains as
mysterious.

\citet{pi05a} mention that their MD based on Washington photometry is
suggestive of a bimodal distribution, with possible peaks around $-$0.8 and $-$1.25.
Our MD also hints at bimodality, with possible peaks around $-$0.9
and $-$1.15, reasonably close to the values suggested by Piatti et al.
The sample size is near the limit of drawing statistically
significant results of this kind.
Indeed, a KMM analysis \citep{ash94} indicates bimodality cannot
be favored over unimodality with any statistical significance.
Clearly, it is of great interest to
enlarge the present sample in order to ascertain whether or not the SMC cluster
MD is indeed bimodal.

\subsection{Metallicity Gradient}

Until recently, essentially nothing was known about the existence or not of
any  variation of metallicity with distance from the center of the SMC. Such metallicity gradients of course
exist in many galaxy disks, including our own, and are  well-traced by the cluster
system \citep[e.g.,][]{jan79}. However, \citet{ols91}, \citet{gei03} and G06 did not find 
any radial metallicity gradient
in the LMC cluster system. \citet{zar94} suggested that disk abundance
gradients are ubiquitous in spiral galaxies, but that the presence of a classical
bar tends to weaken the gradient. The LMC has such a bar, as does the SMC.
Therefore, one might suspect that any metallicity gradient in the SMC may be weak
if Zaritsky et al.'s idea is correct.

Previous searches for any gradient in the SMC have indicated the possible presence of a
weak gradient. \citet{pi07a} found a tendency for the mean cluster metallicity and
its dispersion to be greater inside $4^\circ$ than outside, and \citet{pi07b}
 reinforced this finding. \citet{car08} studied CaT metallicities of a large
number of field giants in each of 13 areas ranging from $1.1 - 3.9^\circ$ and
found evidence for a significant metallicity gradient. 
More specifically, they found that the areas within $\sim2.5^\circ$ have a 
relatively constant mean metallicity of $\sim -1$ but that this value decreases
in their outermost fields, down to $\sim -1.6$ near $4^\circ$.
They noted that this gradient
could be due to either a chemically less-evolved outer region or a similar
chemical evolution (essentially AMR) throughout the galaxy but with a varying mix
of ages, with a larger fraction of younger, more metal-rich stars closer to
the center, or both, and suggested the second scenario is most likely.

In order to investigate any gradient, one must first address the orientation
of the galaxy and projection effects. In the LMC, the orientation of the galaxy
is well-determined and distances can be deprojected to derive true
galactocentric distances. However, the orientation of the SMC is
much more poorly determined and it is also significantly elongated along the
line of sight, making determination of true galactocentric distances much more
problematic. Following \citet{pi07a}, in an effort to derive
galactocentric distances more accurately than simply ignoring projection
effects, we adopted an elliptical coordinate system with $b$/$a$ $=$ 1/2 and
aligned along the Bar, 
as shown in Figure 1. This reference system appears more
appropriate for deriving galactocentric distances than a system simply aligned along
the cardinal directions. We computed for each cluster the value of the semimajor
axis, $a$, that an ellipse would have if it were centered on the SMC center,
aligned with the Bar, had
a $b$/$a$ ratio of 1/2 and one point of its trajectory coincided with the cluster
position, and use this as a surrogate for the true galactocentric distance.
We give the $a$ value so derived for each cluster in Table 3.

In Figure 13 we plot metallicity vs. the semi-major axis $a$ value for our cluster sample and for 
the full       sample. No clear trend is evident. 
Dividing our sample at $4^\circ$, as did \citet{pi07a,pi07b}, we find for
the 15 inner clusters a mean metallicity and standard deviation of -0.94, 0.19,
while the 10 outer clusters have-1.00, 0.21. The difference is not significant.
To check to see if this could possibly be due to an inverse age 
gradient effect, we also checked the mean ages of the two divisions. The inner
clusters are 3.1 (1.9) Gyr and the outer clusters 4.4 (3.4) Gyr. Thus, this
can not be the cause of a lack of an observed metallicity gradient.
We conclude that any
true metallicity gradient in the SMC cluster system must be relatively weak. Note that
our clusters extend to distances twice as far from the center as Carrera et al's field
star sample. It is unclear why they found a gradient and we do not.
Obviously, these intermediate-age clusters and field stars may now occupy a position
in the SMC quite distinct from that where they were born. Additional data are
required to ascertain the existence and strength of any gradient.
If the gradient is indeed minimal, \citet{zar94}'s suggestion that a
strong bar weakens any disk gradient is a viable explanation.

\subsection{Age-Metallicity Relation}

One of our best clues into the detailed chemical evolutionary history of a galaxy
is provided by the relationship between the age of a population and its overall metallicity
and how it varies with time.
This AMR can be compared to models incorporating a variety of variable
parameters concerning the
chemical evolution to test their importance. Clearly, both the ages and metallicities
of the tracer population need to be as accurately determined and on as homogeneous a
scale as possible.

There have been a number of
investigations aimed at deriving the SMC AMR using various stellar populations,
including star clusters, field  giants
and planetary nebulae. The earliest studies by \citet{str85}, \citet{dac91} and  \citet{ols96}
generally agreed on several salient points:
1). There was an initial period of quite rapid enrichment which brought the
metallicity up to $\sim -1.2$, the metallicity (DH98)
of the only old ``globular cluster" in the SMC,
NGC\,121, by the time of its formation some 11 Gyr ago \citep{gl08a}.
Note that this phase is not well traced - it is simply based on the fact that
the oldest cluster was formed from material that had already achieved this level
of enrichment. Indeed, in the latest work with PNe \citep{idi07}, they
find metallicities $\sim -1$ for the oldest planetaries in their sample, with ages approaching
the age of the universe, again demonstrative of the rapidity of this earliest
enrichment phase.
2). For the next $\sim 8$ Gyr, the enrichment was quite slow, although both clusters
\citep{dac91} and field stars \citep{har04} were formed in
significant numbers.
3). Finally, during the past several Gyr, the enrichment has proceeded at a
substantially quicker pace, bringing the global metallicity up to its current value
of $\sim -0.65$ \citep{rus89}.

Thus, prior to the work of DH98, it appeared that the enrichment during the intermediate
age period was very small and that the enrichment in the last few Gyr was
especially dramatic. This sort of AMR is very different from that in the solar
neighborhood as well as that expected from the simple ``closed box" model of
chemical evolution, in which star formation proceeds gradually and hence
abundances change slowly.
In order to account for such effects, one needs to invoke infall of unenriched
gas or outflow of processed gas in a galactic wind, or possibly a burst of star
and cluster formation, which complicate the
chemical evolution.

DH98 obtained the first well constrained AMR of SMC clusters, based on CaT spectra and
deep CMD ages. However, their sample size was small -- only 6 clusters -- so  they
augmented it with data from other sources (as we do).
They observed four intermediate-age clusters and found that the two
oldest, L1 and K3, are slightly more metal rich than the two youngest,
L113 and NGC 339.
They argued that the AMR derived from L\,1 and K\,3 plus their other clusters
was the true global AMR and that L\,113 and NGC\,339 were ``anomalous" and
required an additional explanation, such as infall of metal-poor gas. The AMR
derived from the other objects yielded a smooth, continuous enrichment
throughout the history of the galaxy, with no change in the rate of increase in
the last few Gyr. According to them, the metallicity increased from $\sim -1.2$ to
$\sim -0.75$ between $\sim 10 - 3$ Gyr, and continued to increase at a slightly
smaller rate in the last 3 Gyr. They found that a simple closed box model
(shown in Figure 14 as the short dashed curve) fit their
data well (with the exception of the 2 problematic clusters) and that neither
infall nor outflow of gas nor any discontinuous process
was required for the global trend.
However, note that their conclusions
would change substantially if they instead considered L\,1 and K\,3 as anomalous,
in which case their AMR
would resemble much more those of previous
studies, with a much slower enrichment during the intermediate
period.

Subsequent AMR investigations have tended to corroborate the original
studies and reinforce the need for something more complicated than the
simple model proposed by DH98. In a series of papers investigating
cluster ages and abundances from Washington photometry, currently culminating in
\citet{pi07c}, Piatti and collaborators compiled this
information for a total of almost 50 clusters. Their AMR clearly shows very little enrichment in the intermediate
period and substantial enrichment beginning about 3 Gyr ago. This result
is in excellent agreement with the model of \citet[hereafter PT98]{pag98} which assumes an early burst, followed by no star
formation from $\sim12 - 4$ Gyr ago, and then an ensuing burst of recent
star formation, leading to a significantly enhanced subsequent
enrichment. The PT98 AMR is shown in Figure 14 as the solid curve. 
Using the $UBVI$ photometry from their Magellanic Clouds Photometric
Survey, \citet{har04} derived the AMR for both field stars
and clusters in the SMC and found the two AMRs to be consistent with
each other.  Their final relation, shown in Fig.~14 as a dotted line, is
similar to the PT98 AMR for younger ages but has substantially enhanced
metallicities for ages $>2$ Gyr.  Note that the \citet{har04} 
study was not sensitive to objects older than 10 Gyr. Based on 29 PNe,
\citet{idi07}  built an AMR that is similar to that of Harris \&
Zaritsky but offset to even higher metallicities in the intermediate age
range by $\sim 0.2$ dex and shows virtually no change in metallicity
during this period.  \citet{noe07} determined the AMR for a field
$1.1^\circ$ to the southeast of the SMC center using $B-$ and $R-$ band
photometry and a detailed analysis including isochrones and the star
formation history.
Their AMR is corroborated in an independent study by \citet{car08}, which includes some fields in common with \citet{noe07}. 
\citet{car08} used the CaT to derive metallicities on the CG97
scale for a large number of field giants in regions spread across the
SMC. They find the AMRs for each region are indistinguishable and that
there is thus a global AMR. We plot the mean age and metallicity
points from \citet{car08} in Fig.~14 (open squares) along with
the derived best-fit model (long dashed curve).  The model, which comes
from \citet{car05}, did not assume any bursts but instead used both
infall and outflow.

Our clusters are presented in Figure 14, along with the additional  samples
described above. A summary of the ages and metallicities that we are considering 
for these samples, as well the corresponding references, is given in Table 5. 
The overall agreement with the PT98  model is
good, albeit there are two     regions of disagreement. 
The predicted metallicity is generally lower than observed from $5 - 10$Gyr, 
and our 4 youngest clusters as well as NGC 330 are
significantly more metal-poor than the PT98 prediction. DH98's 2 ``anomalous"
clusters (which have exactly the same assigned age and metallicity)
are now very well behaved with respect to the PT98 model
and in fact it is L\,1 and K\,3 which appear
anomalous (too metal-rich)! Note that the ages of all of the DH98 clusters have
changed in \citet{gl08b}'s derivation and should be much better determined
now. We therefore adopt the ages of Glatt et
al., which are derived via main sequence fitting with the Dartmouth
isochrones.
As pointed out by \citet{gl08b}, a small age gap in the cluster 
distribution now appears between the oldst SMC cluster (NGC 121 $=$10.5 Gyr)
and the second oldest cluster (L1 at 7.5 Gyr). This is much briefer than the
infamous LMC cluster age gap, where only 1 cluster is found to have an age
between $12 - 3$Gyr. Curiously, this lone LMC age-gap cluster was formed 
in the middle of the SMC ``age gap''. In any case, the SMC gap appears real and is
interesting, and no subsequent gaps of more than a Gyr appear.
The very minimal chemical enrichment predicted by PT98 in the intermediate
period and the fast and continuing rise starting about 3 Gyr ago are
particularly well reproduced by the data (with the exception of the youngest
clusters). Note that L\,11, the DH98 point at (3.5 Gyr, $-$0.75),
 has a CaT value estimated by \citet{kay06} to be $\sim -0.9$,
which would bring it more in line with the PT98 prediction. 
The Carrera model gives a similarly good overall fit, especially in
the two regimes where the PT98 model is poorest.

Thus, our data are in good overall
agreement with the PT98 burst model, as well as with
the hybrid infall $+$ outflow model of \citet{car05}. The ideal model would 
predict metallicities between these 2 models. It is clear that
something more complex than the simple closed box model of DH98 is required to
explain the observations: the rate of enrichment during the intermediate period was
certainly much less than over the last few Gyr. 
In fact, the metallicity appears to have stayed almost constant at $\sim -1.1$
from $10 - 3$Gyr and then had a later increase. Our data are also in very good
agreement with Carrera et al.'s field star data, indicating, as they first 
discovered, that the AMR was universal.
We will focus our discussion on the burst model.

The possibility of a burst of star formation to explain the AMR in the SMC
was first suggested by \citet{tsu95} and developed by PT98.
They assumed a burst at 4 Gyr, partly based on the AMR's available to them.
Note that their model includes infall of
unprocessed material as well. A  problem with their burst model is that they predict
no star formation from 12 $-$4 Gyr, although it is clear this was NOT the case
and that an important fraction of both the stellar and cluster populations were formed
during this period \citep[e.g.,][]{mcc05,dac91}.
\citet{ric00} argued from HST imaging for 7 clusters that there were 2 main
epochs of cluster formation: one at $8\pm2$ Gyr and another at $2\pm0.5$ Gyr, and that
the metallicities of the clusters in each group are also very similar. 
\citet{pi05a,pi07b},
from a much larger sample of clusters but with less precise age determinations,
suggested possible bursts at about these same times. Note that such bursts
could possibly explain the metallicity bimodality if it is real.
\citet{har04} found peaks in the age distribution at 0.4 and 2.5 Gyr in the
global field star formation history but no significant peaks in the cluster age distribution.
\citet{idi07} also claimed their AMR supported a burst that occurred
in the last 2 -- 3 Gyr and that there was very little chemical evolution before
this time going back to $>10$ Gyr ago.

The case for reduced star formation and chemical enrichment during the intermediate
period from $\sim 10-3$ Gyr and a recent period of enhanced star formation and enrichment over the last
few Gyr is well supported by a preponderance of the data currently available,
including our own. The question obviously arises as to what caused the recent
burst? \citet{bek04} analyzed the orbits of the Magellanic Clouds and
concluded that the first close encounter between them occured about 4 Gyr ago
and this could have led to strong triggering of star and cluster formation.
Unfortunately, our current knowledge of the Clouds' proper motions prohibits
us from making definitive statements or prediction, and indeed
the most recent values all suggest the Magellanic Clouds may be
unbound to each other and only beginning their first close encounter to the MW 
\citep{kal06,pia08}. Note that the LMC also experienced a recent burst
at approximately the same time but clusters born over the last few Gyr show a
very uniform metallicity, with no signs of the metallicity increase that most of their SMC
cousins underwent. G06 find that their sample of 23 LMC intermediate-age 
clusters (1 -- 3 Gyr old) have a very tight metallicity distribution, with a
mean of -0.48 and a standard deviation of only 0.09 dex. In our SMC sample,
the 10 clusters in the same age range have a mean of -0.82 and $\sigma = 0.15$,
and thus are both significantly more metal-poor and with a wider metallicity 
spread than their LMC counterparts. Clearly, more high quality ages and metallicities,
as well as detailed elemental abundances, for SMC clusters are required to help
constrain its chemical evolution.

\section{Kinematics}

The structure of the SMC is known to be extremely
complex, with a large line-of-sight depth that may vary across the face
of the SMC \citep[e.g.,][]{cro01}, possibly a result of interactions
with the LMC and Galaxy.  Some studies of the SMC's kinematics suggest
that they may be equally complex.  For example, \citet{sta04} used \ion{H}{1} 
observations to examine the kinematics of the SMC and
found what appears to be differental rotation, with a turnover radius of
$\sim$ 3 kpc.  The authors suggest that the central region of the SMC
may correspond to a disk-like structure left over from when the SMC was
rotationally supported, a time prior to any interactions with the LMC and
MW.

Other authors, however, did not find evidence of rotation. \citet{har06} summarized 
previous radial velocity
studies of the SMC. As seen in their Table 3, these are based on
the study of a variety of different objects, including planetary nebulae
\citep{dop85}, red giant stars \citep{sun86,har06}, supergiants and main sequence stars 
\citep{mau87}, cepheids \citep{mat88}, carbon stars \citep{har89,hat97}, red clump stars 
\citep{hat93} and expanding \ion{H}{1} shells \citep{sta97}.  
The mean heliocentric radial velocities (ranging from 123 to 162 km
s$^{-1}$) and velocity dispersions (ranging from 18 to 33 km s$^{-1}$)
are in good agreement, regardless of the type of tracer
used in the analysis.
From analysis of their own data (2046 SMC stars in 16 fields), \citet{har06} found a velocity 
gradient that may include a signature of rotation in the SMC. They found an apparent rotation amplitude of 8.3
km s$^{-1}$ deg$^{-1}$ and a rms scatter of 27.5 $\pm$ 0.5 km s$^{-1}$.

More recently, \citet{car05}, obtained similar
results with stars in 13 fields of the SMC. He observed that the velocities are compatible with the
presence of a small rotation, with an amplitude of $30\pm4$ km s$^{-1}$
and maximum radial velocity at PA$\sim 150^{\circ}$.  However, both Carrera 
and Harris \& Zaritsky, noted that part or
all of this ``rotation" could in fact be due to the projected tangential
motion of the galaxy. Given that the possible rotation velocity
is much smaller than the velocity dispersion, both concluded that the SMC
is primarily supported by its velocity dispersion.

Regarding star clusters, the only previously published work is that of
DH98, who used their CaT spectra to obtain radial
velocities of seven SMC clusters. They found a mean radial velocity of
138 $\pm$ 6 km s$^{-1}$ and a velocity dispersion of 16 $\pm$ 4 km
s$^{-1}$, values in reasonable accord with those summarized above.  From
their analysis, they support the conclusion of the lack of evidence for
any systematic rotation of the SMC.

To test for global rotation, we have analized the behaviour of the
cluster radial velocities derived by us as a function of position angle.
The velocity data of our sample are plotted in Figure 15 (solid circles).
To calculate the position
angle we have used the equations of \citet{van01} which
allow us to perform the conversion from right ascension and declination
to Cartesian coordinates using a zenithal equidistant projection. We
adopted the SMC optical center given by \citet{cro01},
$\alpha$ = 00$^h$ 52$^m$ 45$^s$ and $\delta$ = 72$^{\circ}$ 49' 43"
[J2000.0] and the usual astronomical convention that position angle is 0
to the north and increases to the east. We have also
included in the graph the seven clusters from DH98 (open circles ).

Applying to our sample the maximum likelihood nethod (explained, for example, in the appendix B of \citealt{har94}), we
have calculated a mean heliocentric velocity of 148 $\pm$ 6 km s$^{-1}$ and a
velocity dispersion of 23.6 $^{+5.7}_{-3.3}$ km s$^{-1}$, in agreement with the
DH98 cluster results.
 When we include in the calculation the
DH98 clusters, the mean heliocentric velocity and the velocity
dispersion are 145 $\pm$ 5 and 22.3 $^{+4.2}_{-2.6}$ km s$^{-1}$ respectively.  These values
compare well with previously derived values, suggesting
that the star cluster system shares the same kinematics as other tracers
in the SMC. 

The distribution of the points in Figure 15 may hint at rotation with a
maximum at position angle of $\sim$ 90 degrees and an amplitude of perhaps 20 km s$^{-1}$.
However, this result is not statistically significant, as the fitting of the points
(curve in Figure 15) is not statistically better than a straight line.
 If any apparent rotation is indeed present, we must agree  with \citet{har06} that this 
could be due to projected tangential motion and that any possible
 rotation velocity is smaller than the velocity dispersion.

\section{Summary and Conclusions}

Magellanic Cloud clusters are an excellent laboratory for helping to
unlock the secrets
of cluster and galaxy formation. They are also crucial testbeds for stellar 
and chemical evolution
models and interpreting the integrated light of distant galaxies. However,
despite their utility and proximity, Small Magellanic Cloud clusters in particular
have been overlooked in this regard. In order to help remedy this situation,
we have carried out a large-scale investigation of the kinematics and
metallicities for a number of SMC star clusters. Building on our experience with
deriving these parameters for LMC clusters using the CaT technique (G06), we
have used the same technique to explore SMC clusters. We obtained CaT spectra
for a number of stars associated with 16 SMC clusters using the VLT + FORS2
MXU instrument. This provides the largest sample of SMC clusters with
velocities and well-determined spectroscopic metallicities currently available,
more than doubling the only previous study (DH98). Target clusters were selected 
from our Washington photometric system
studies, which provide a rough estimate of age and abundance, to be old
enough to possess red giant branch stars and cover a wide area across the
galaxy.

We used the same reduction and analysis techniques as we did in G06 to
measure stellar radial velocities  and metallicity. Typical
radial velocities errors are 7.5 km s$^{-1}$ and metallicity errors are 0.17 dex per star. Cluster
members are selected using an analysis combining their location in the
CMD, position in the cluster, and radial velocities and metallicity with respect to other stars in the same
field. We determine mean cluster velocities to
typically 2.7 km s$^{-1}$ and metallicities to 0.05 dex (random error), from a mean of
6.4 members per cluster.  A comparison of our mean cluster metallicities with those
derived from Washington photometry for 11 clusters in common shows very good overall
agreement, with the CaT metallicities being much more precise. This indicates
that the Washington metallicities,  and especially their
procedure used to correct the age dependence of their metallicities for their age dependence,
is appropriate.

The metallicity distribution (MD), metallicity gradient and age-metallicity relation (AMR)
are investigated,
combining our clusters with those observed by DH98 and Glatt et al.
(also with CaT) and the one cluster with a detailed, high resolution metallicity
to compile a sample of 25 clusters on a homogenous metallicity scale with 
relatively small errors,
although the ages are somewhat heterogeneous. The mean metallicity is [Fe/H] $=-0.96$, with
$\sigma = 0.19$. Dividing the sample into two age groups at 3 Gyr, the 12 older
clusters have a mean metallicity of $-1.08, \sigma = 0.17$, while the 13 younger have $-0.85,
0.15$. 
 Most clusters lie between [Fe/H] = $-$0.75 and $-$1.25. There is a suggestion for
bimodality in the MD, with peaks at [Fe/H] $\sim -0.9$ and $-$1.15. No clear
gradient is seen with distance from the center of the SMC. 
However, intermediate-age SMC clusters are both significantly more metal-poor
and have a larger metallicity spread than their LMC counterparts. The AMR shows
evidence for 3 phases: a very early ($>11$ Gyr) phase in which the metallicity reached
$\sim -1.2$, a long intermediate phase from $\sim 10-3$ Gyr in which the metallicity
only slightly increased although a number of clusters formed,
and a final phase from 3-1 Gyr ago in which the rate of
enrichment was substantially faster. These salient features agree with those
found by most other AMR studies using other tracers and techniques.
We find good overall
agreement with the model of PT98 which assumes a burst of star formation at 4 Gyr.
A hybrid infall $+$ outflow model of \citet{car05} also  fits the data reasonably
well. The simple
closed box model of DH98 yields a much poorer fit, and the AMRs derived by \citet{har04}  and 
\citet{idi07}  are
significantly offset to higher metallicities for intermediate-age clusters.
A number of different lines of
evidence point to the likelihood of a burst in the SMC star and cluster
formation about 3 Gyr
ago. The cause of such a burst is currently a source of much speculation. The
suggestion by \citet{bek04} that it is due to a close passage of the SMC
and LMC is intriguing but requires better knowledge of their orbits, especially
proper motions, to be definitively tested.

We finally examine the kinematics of our CaT clusters. Their mean radial velocity is = 148 km s$^{-1}$,
with a velocity dispersion of 23.6 km s$^{-1}$. These values are in very good agreement
with those found by DH98 from their smaller sample. Combining the 2 cluster samples,
we find
the kinematics are dominated by the velocity dispersion, as found in
virtually all other kinematic studies of a wide variety of SMC populations.\\

\acknowledgments

This work is based on observations collected at the European Southern Observatory,
Chile, under program number 076.B-0533. We would like to thank the 
Paranal Science Operations Staff. We would like to thank to the referee for comments that helped to improve this work.
D.G. gratefully acknowledges support from the Chilean
Centro de Astrof\'\i sica FONDAP No. 15010003
and the Chilean Centro de Excelencia en Astrof\'\i sica
y Tecnolog{\'\i}as Afines (CATA). M.C.P. acknowledges Andrew Cole and Andr\'es Piatti for their explanations
and discussion about how to perform some calculations and Ricardo Carrera
for allowing us to use his unpublished age-metallicity models. M.C.P. and J.J.C. gratefully acknowledge 
financial support from the Argentinian 
institutions CONICET and SECYT (Universidad Nacional de C\'ordoba).

{\it Facilities:} \facility{VLT: Antu (FORS2)}.

\begin{figure}
\plotone{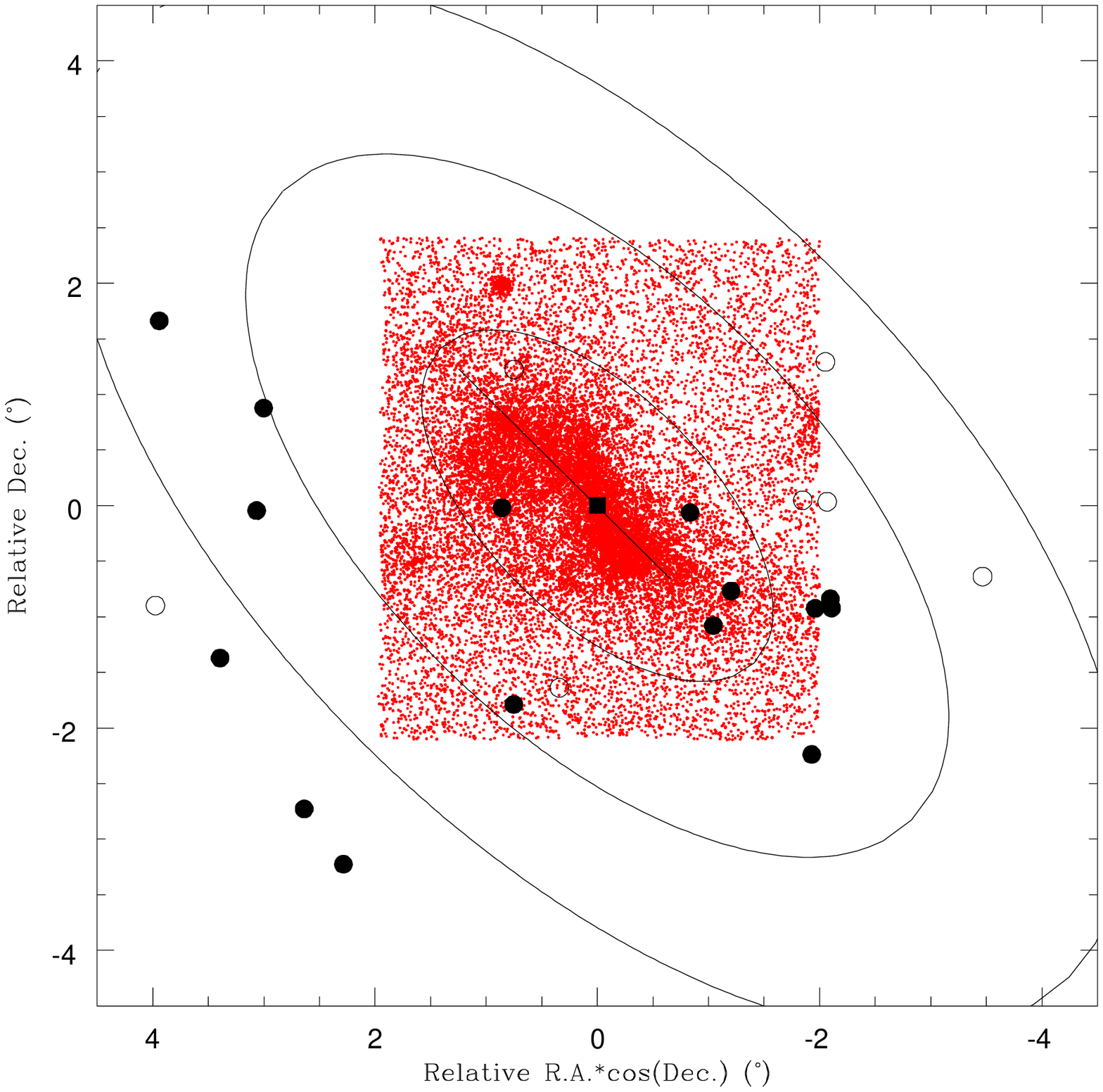}
\caption{Position of our target clusters (filled circles) in relation to the SMC 
optical centre (square) and the SMC bar (line). Open circles represent the 
position of clusters from DH98. The ellipses are aligned with the bar and have 
semi-major axis of 2, 4 and 6 degrees. Following the idea of \citet{gl08b}, a star map 
of the SMC is superposed. This map was generated by using the catalog 
of the Small Magellanic Cloud Photometric Survey \citep{zar02} for 
stars with V $<$ 16. 
}
\end{figure}

\begin{figure}
\plotone{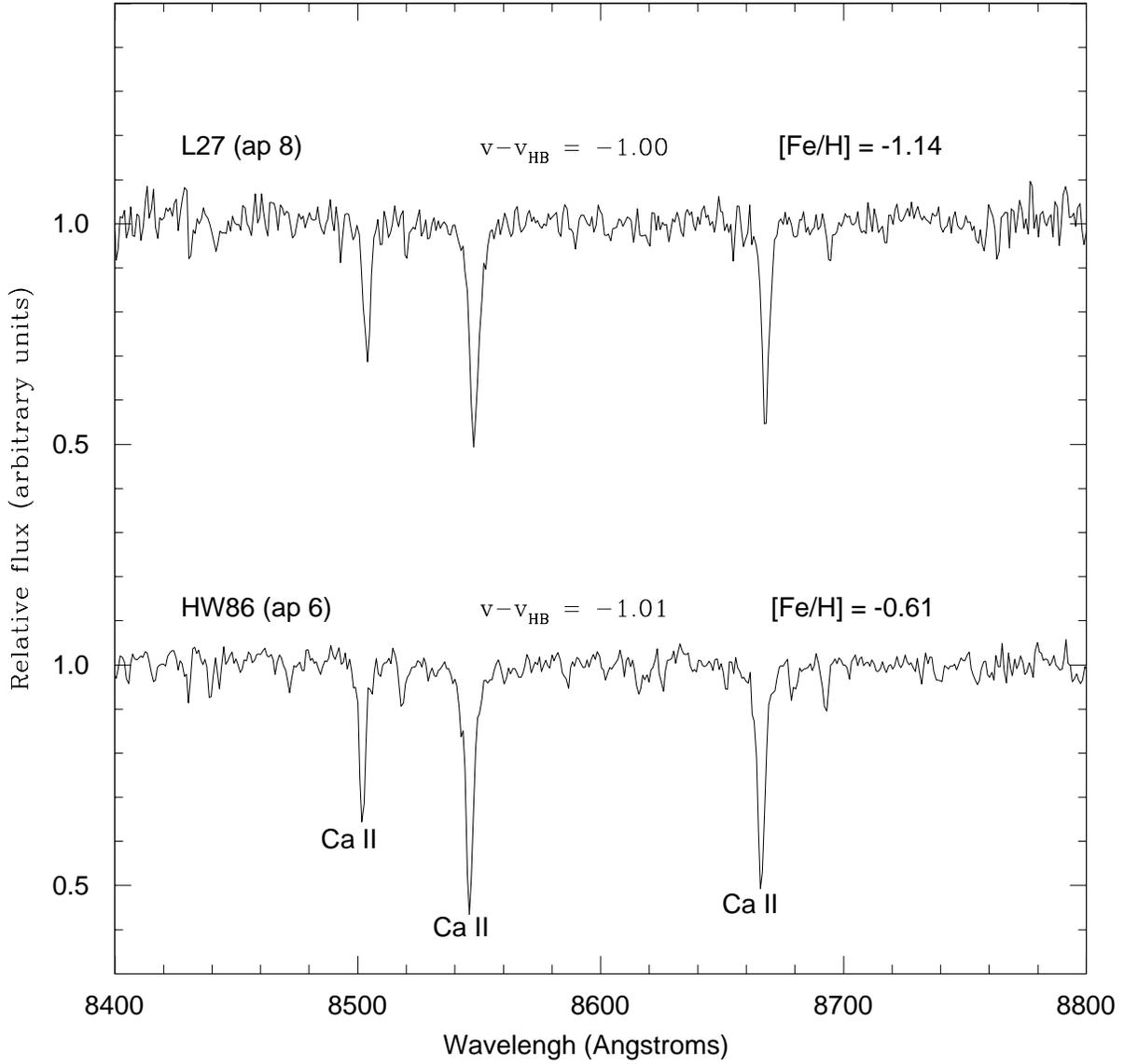}
\caption{Sample continuum-normalised spectra of   RGB stars in two clusters of our sample. The three 
CaT lines have been marked on the plot as well as the corresponding $v-v_{HB}$ 
values and metallicities. These 2 stars have very similar $T_{effs}$ and log g
values. Thus, the clear difference in Ca II line strength illustrates their
substantial metallicity difference. }
\end{figure}

\begin{figure}
\plotone{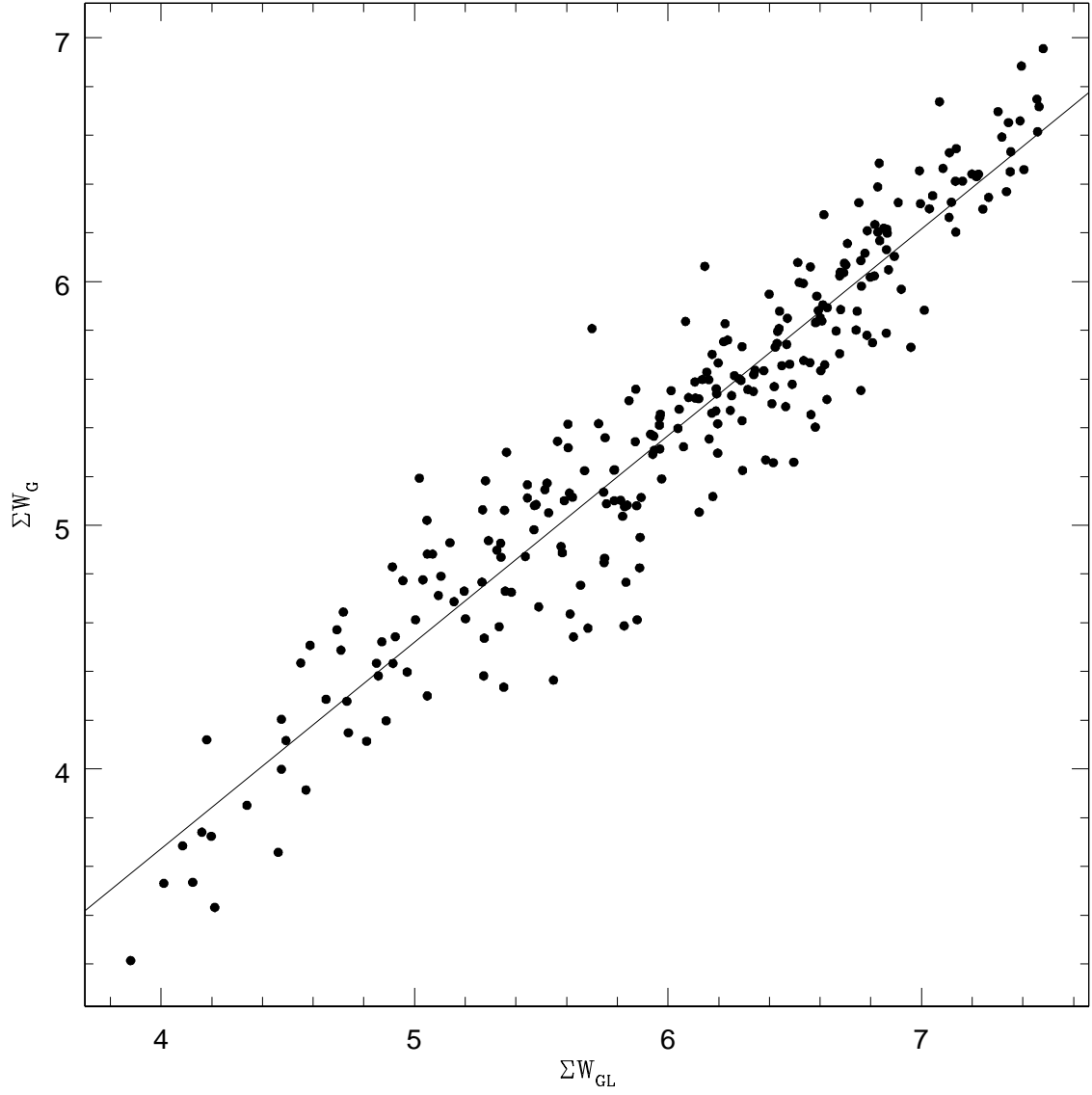}
\caption{Relation between the Gaussian-fit EW, $\Sigma W_G$, and the
Gaussian $+$ Lorentzian-fit, $\Sigma W_{GL}$. The points correspond to
spectra with S/N $>$ 20 and the solid line shows the linear fit to the data.}
\end{figure}

\begin{figure}
\plotone{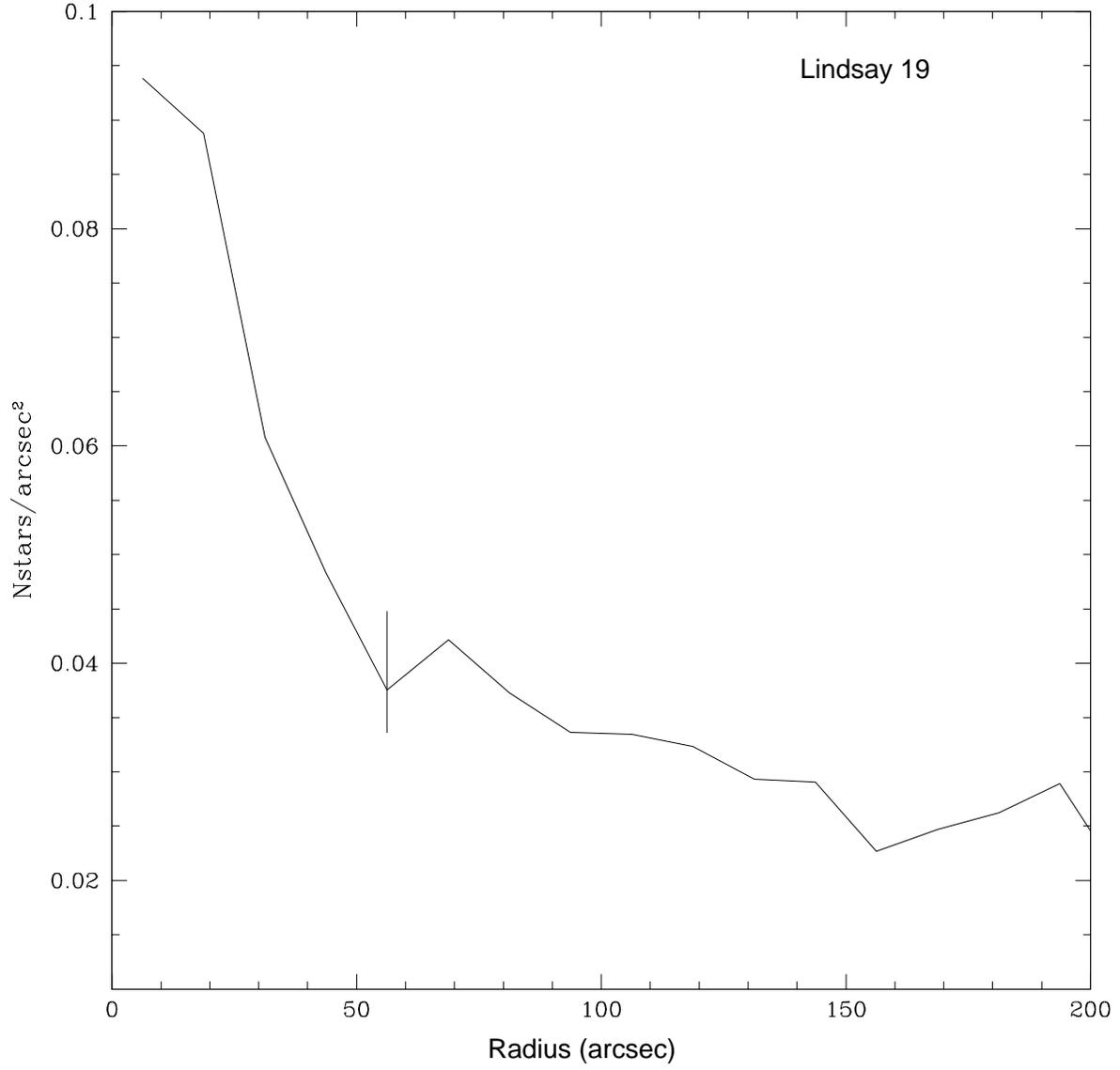}
\caption{Radial stellar density profile of the cluster Lindsay\,19. The x-axis represents 
the radius and the y-axis represents the stellar number density. The adopted cluster
radius is marked.}
\end{figure}

\begin{figure}
\plotone{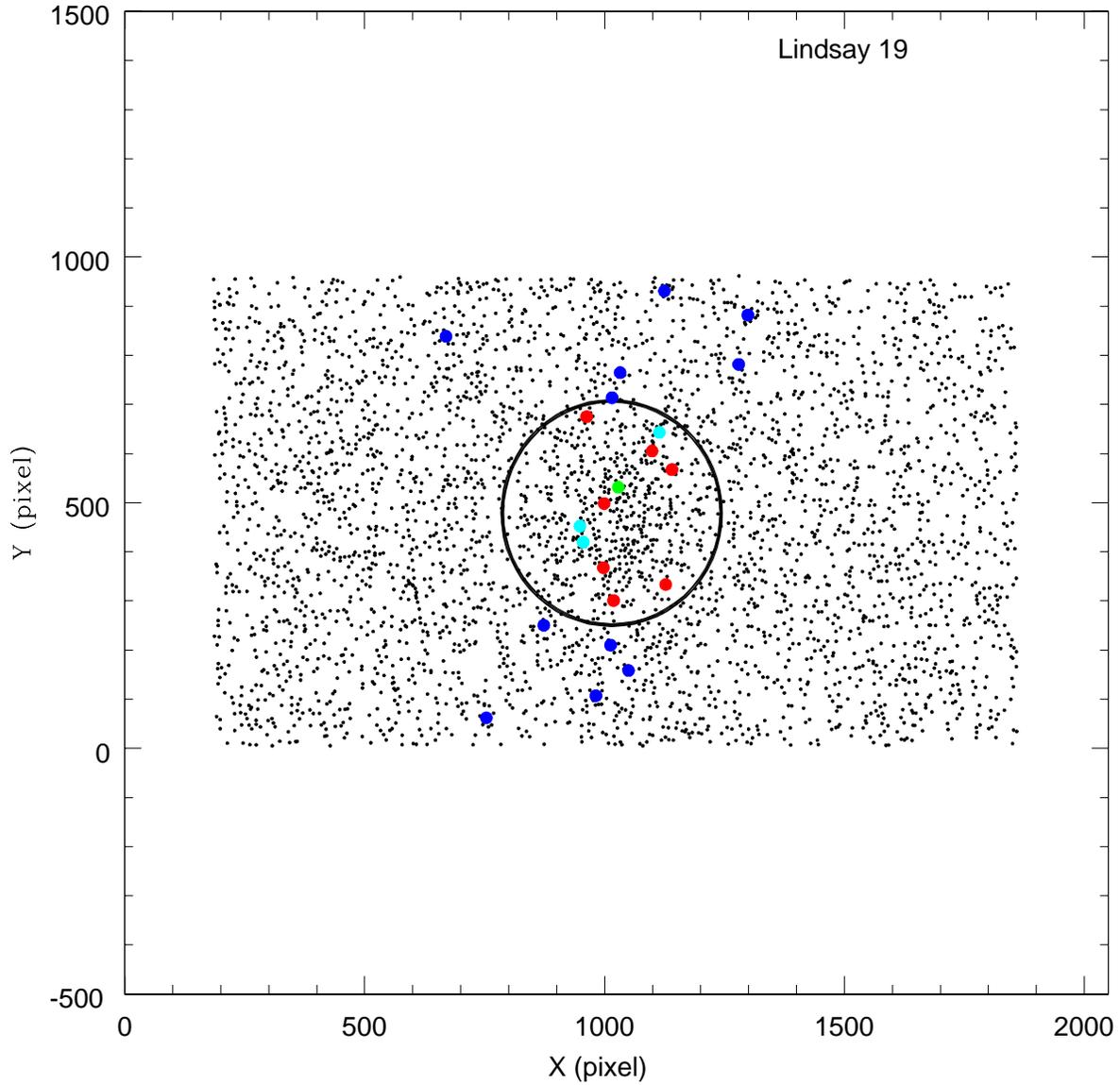}
\caption{Chart of the cluster Lindsay\,19. Our spectroscopic target stars are represented 
by the large filled circles and the adopted cluster radius is represented by a large 
circle. Color code: non-members that are outside the cluster radius (blue circles),
 non-members that were eliminated because of discrepant radial velocity or  metallicity 
(teal and green circles, respectively) and final cluster members (red circles).}
\end{figure}

\begin{figure}
\plotone{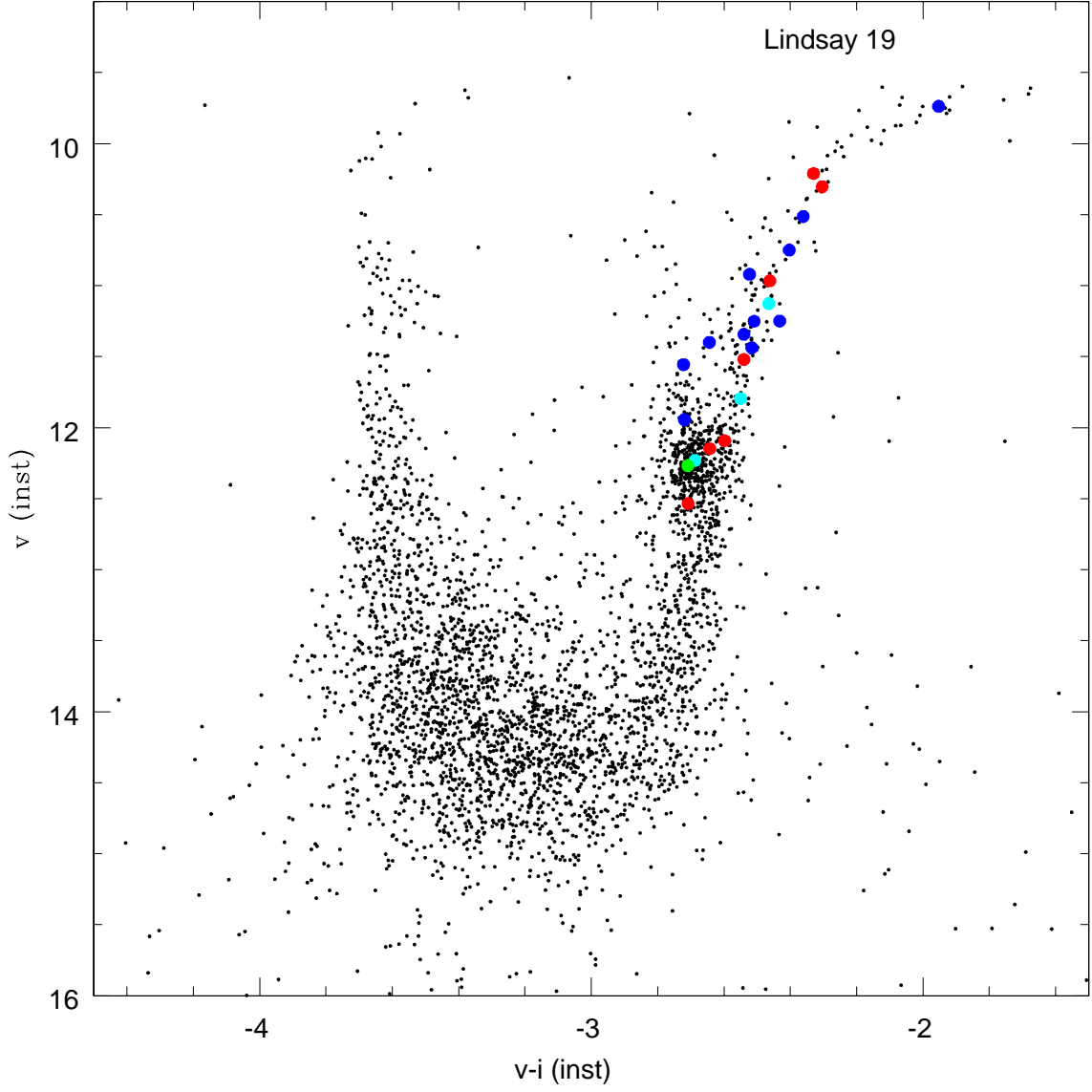}
\caption{Instrumental color-magnitude diagram of Lindsay\,19. The color code is the 
same as in Figure 5.}
\end{figure}

\begin{figure}
\plotone{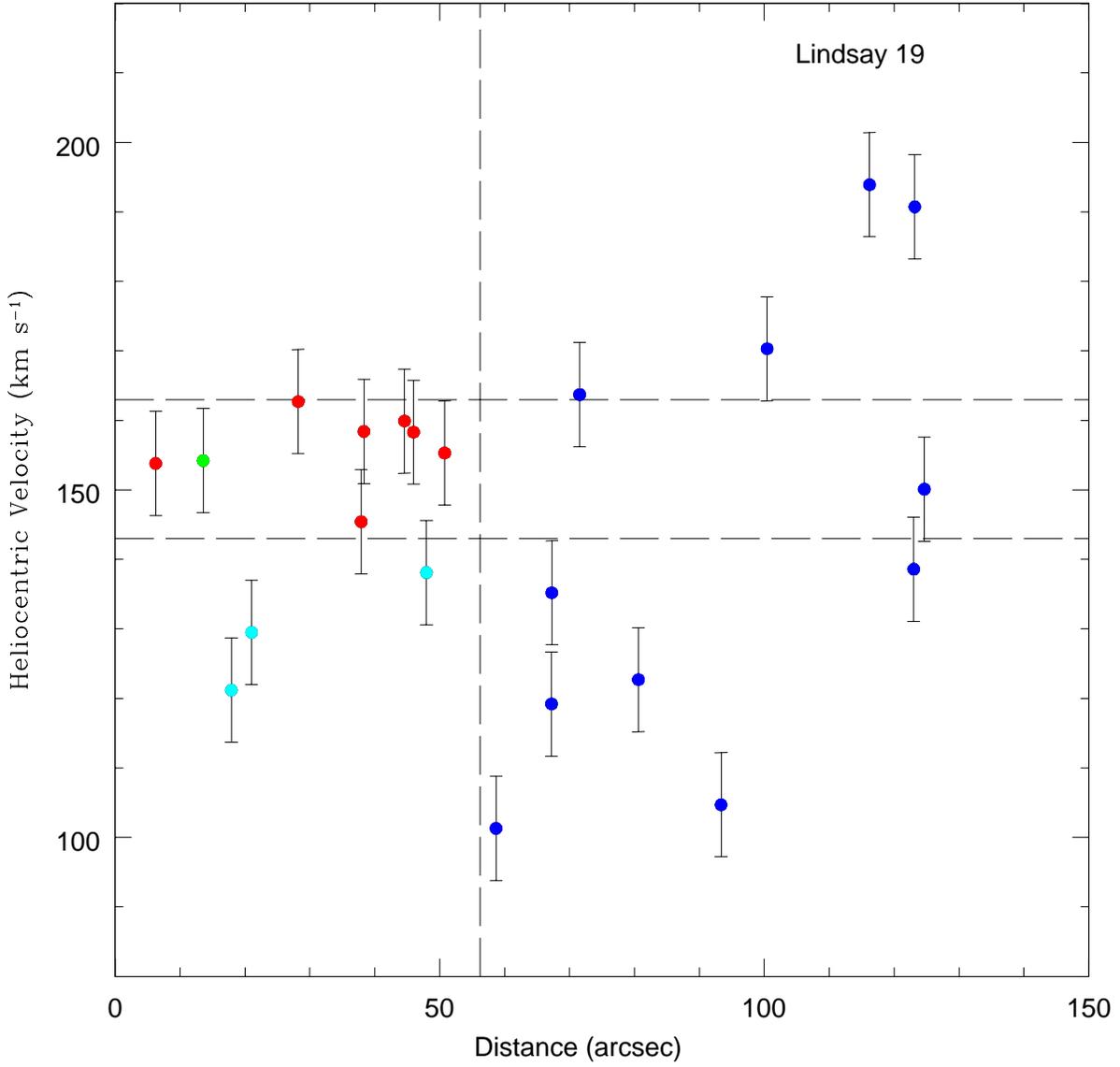}
\caption{Radial velocity  vs. distance from the cluster center for Lindsay\,19 targets.
The horizontal lines represent our velocity error cut ($\pm$10 km s$^{-1}$)
 and the vertical line represents the adopted cluster radius. The color code is the 
same as in Figure 5. Error bars represent the random error in determining the RV for each star.}
\end{figure}

\begin{figure}
\plotone{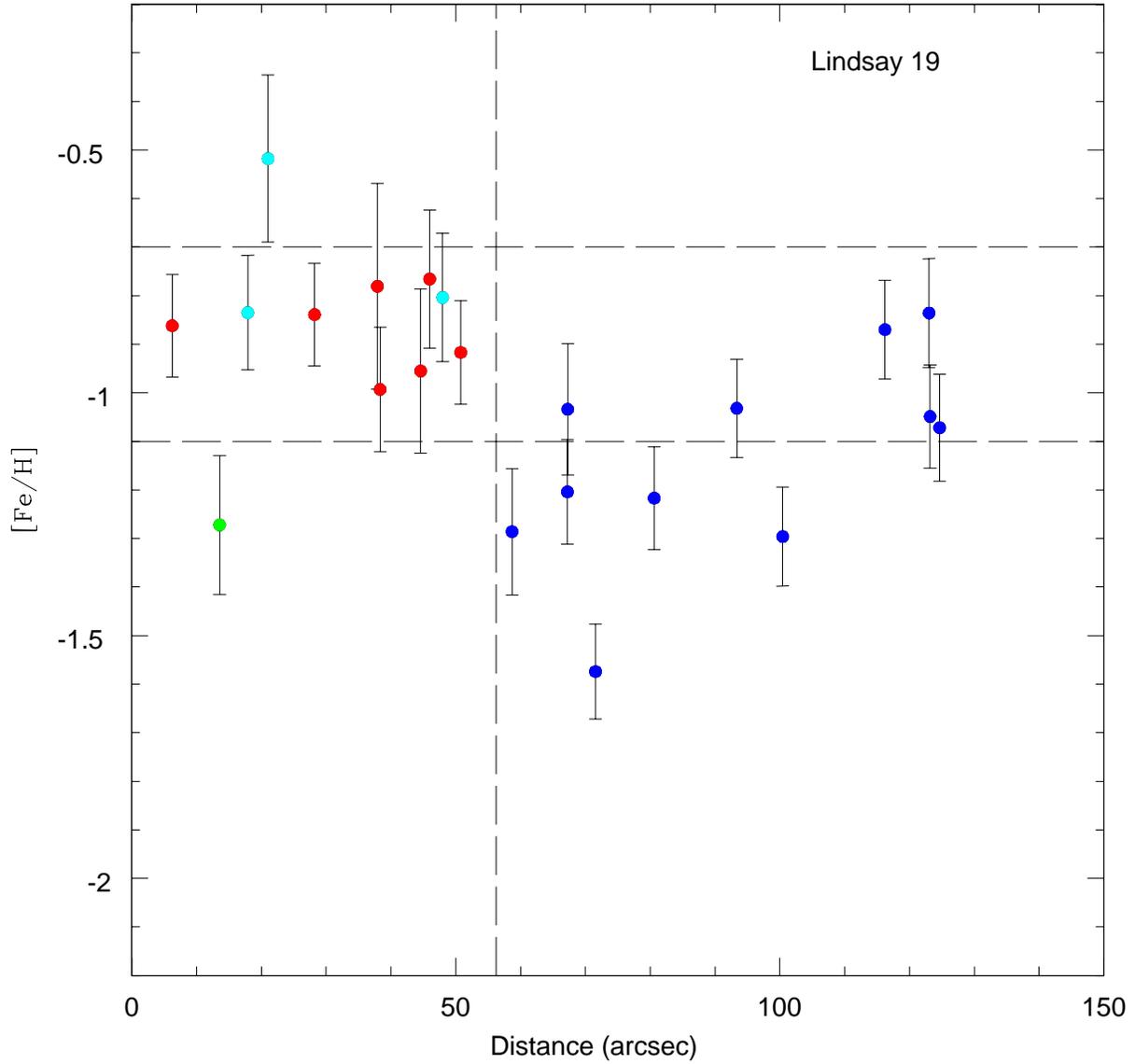}
\caption{Metallicity versus distance from the cluster center for Lindsay\,19
targets. Horizontal lines represent the [Fe/H] error cut ($\pm$ 0.20 dex ) and the
vertical line represents the adopted cluster radius.  The color code is the same as in 
Figure 5. Error bars represent the random error in calculating [Fe/H].}
\end{figure}

\begin{figure}
\plotone{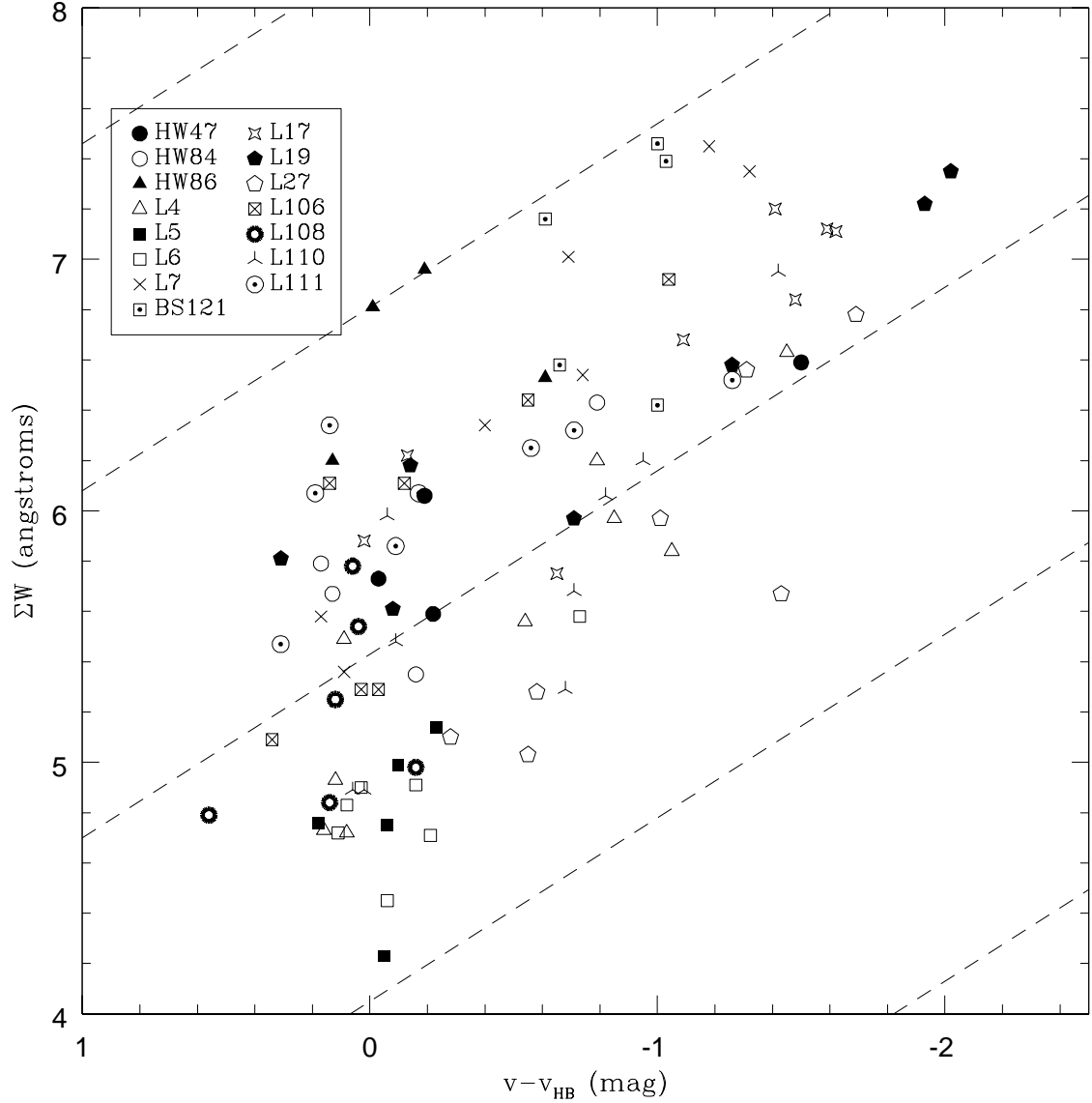}
\caption{The sum of the equivalent width of the three CaT lines versus $v-v_{HB}$ for 
identified members of all clusters.  The dashed lines represent isometallicity lines of 0, $-$0.5, $-$1, $-$1.5 
and $-$2 (from top to bottom).}
\end{figure}

\begin{figure}
\plotone{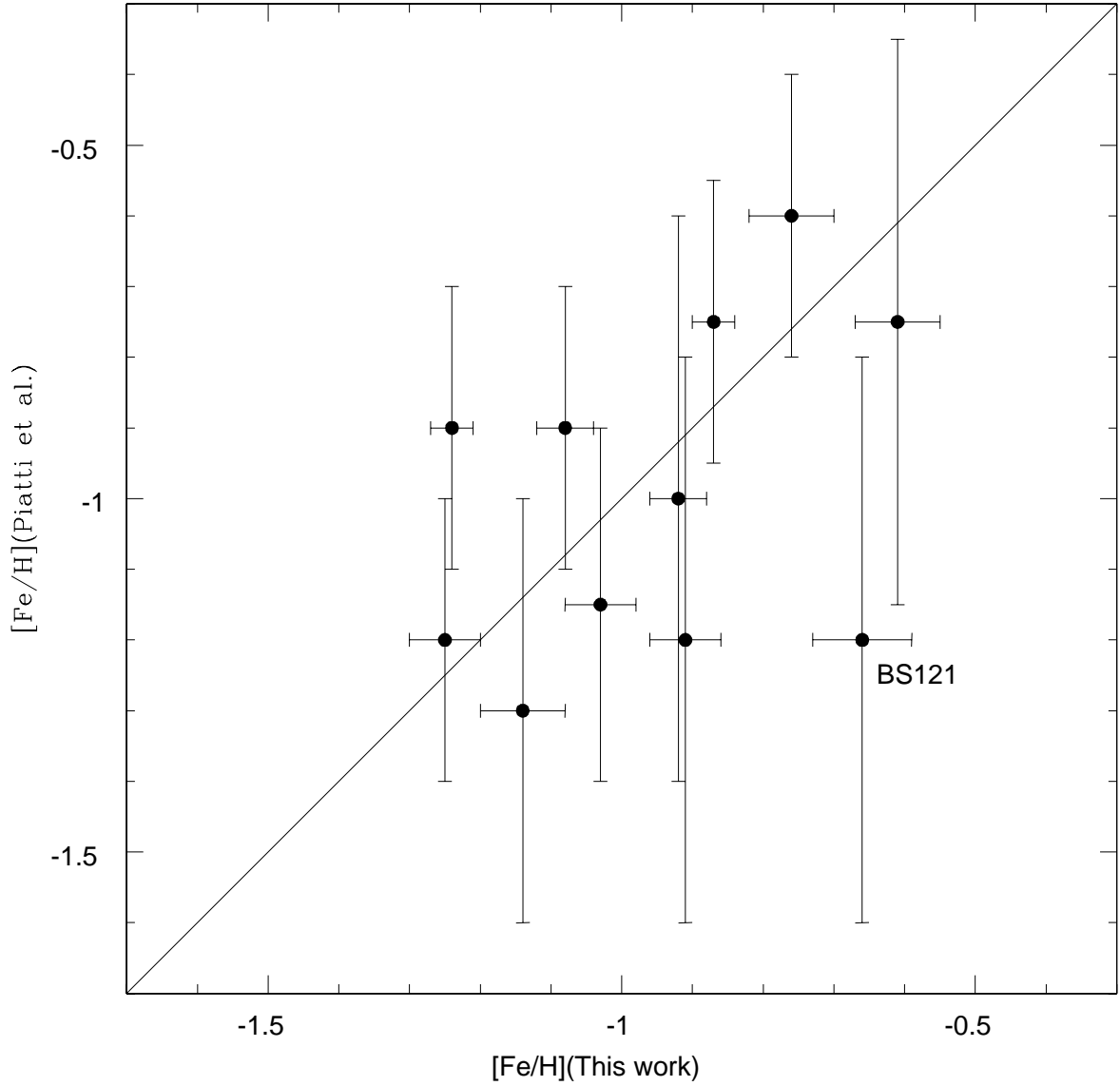}
\caption{Comparison of our spectroscopic mean  cluster metallicities and those derived 
from Washington photometry by \citet{pi05a,pi07b}. The line shows one-to-one
correspondence.
Note that there is no systematic difference between the two systems but that
the error of the CaT technique is substantially smaller. The only cluster showing a 
considerable difference between both metallicity determinations, BS\,121, is marked. }
\end{figure}

\begin{figure}
\plotone{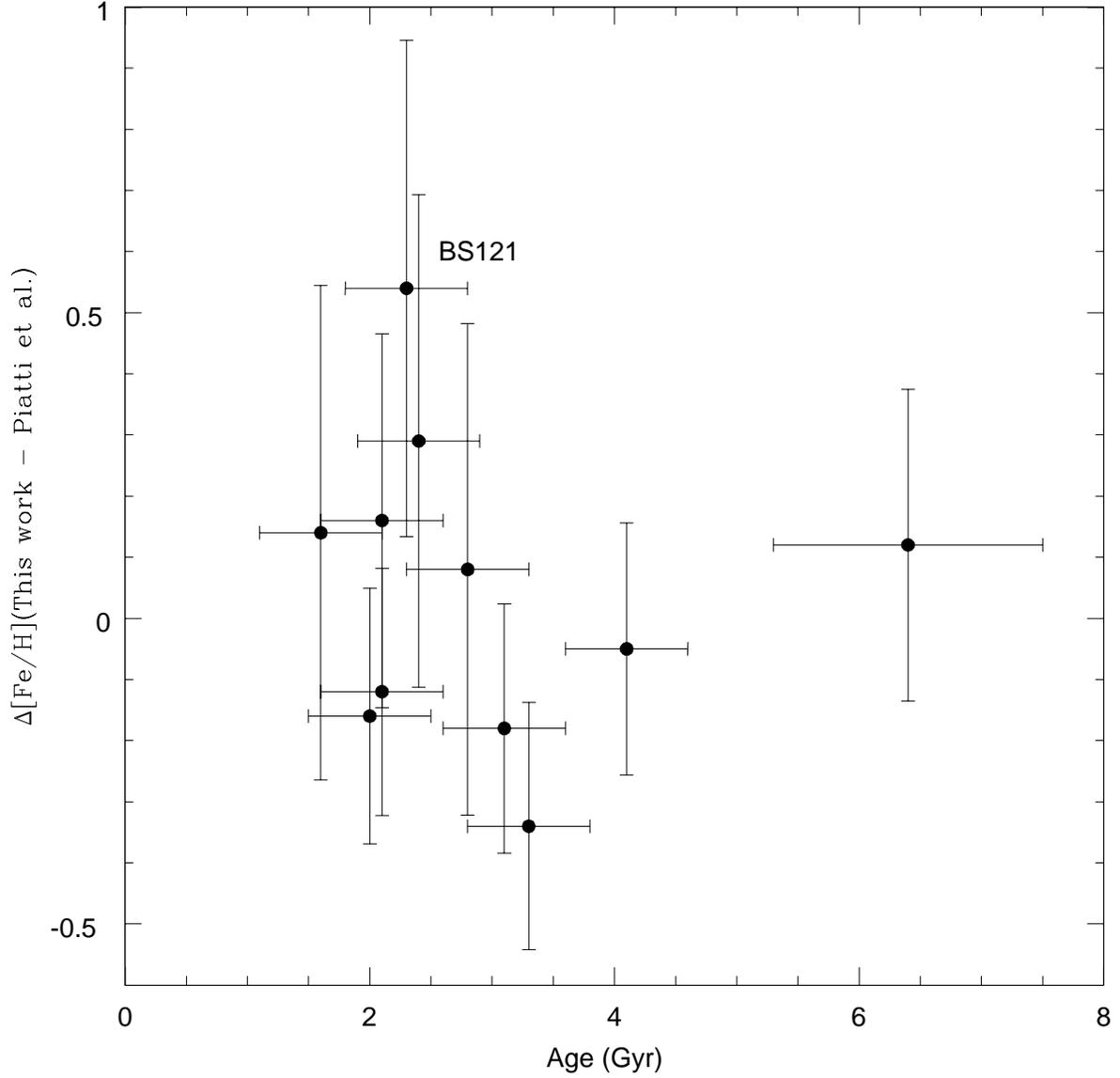}
\caption{Difference between our spectroscopic mean cluster
metallicity and that  derived from Washington photometry by
\citet{pi05a,pi07b} vs. age.
No systematic trend is observed. The only cluster showing a considerable
difference between both metallicity determinations, BS\,121, is marked.}
\end{figure}

\begin{figure}
\plotone{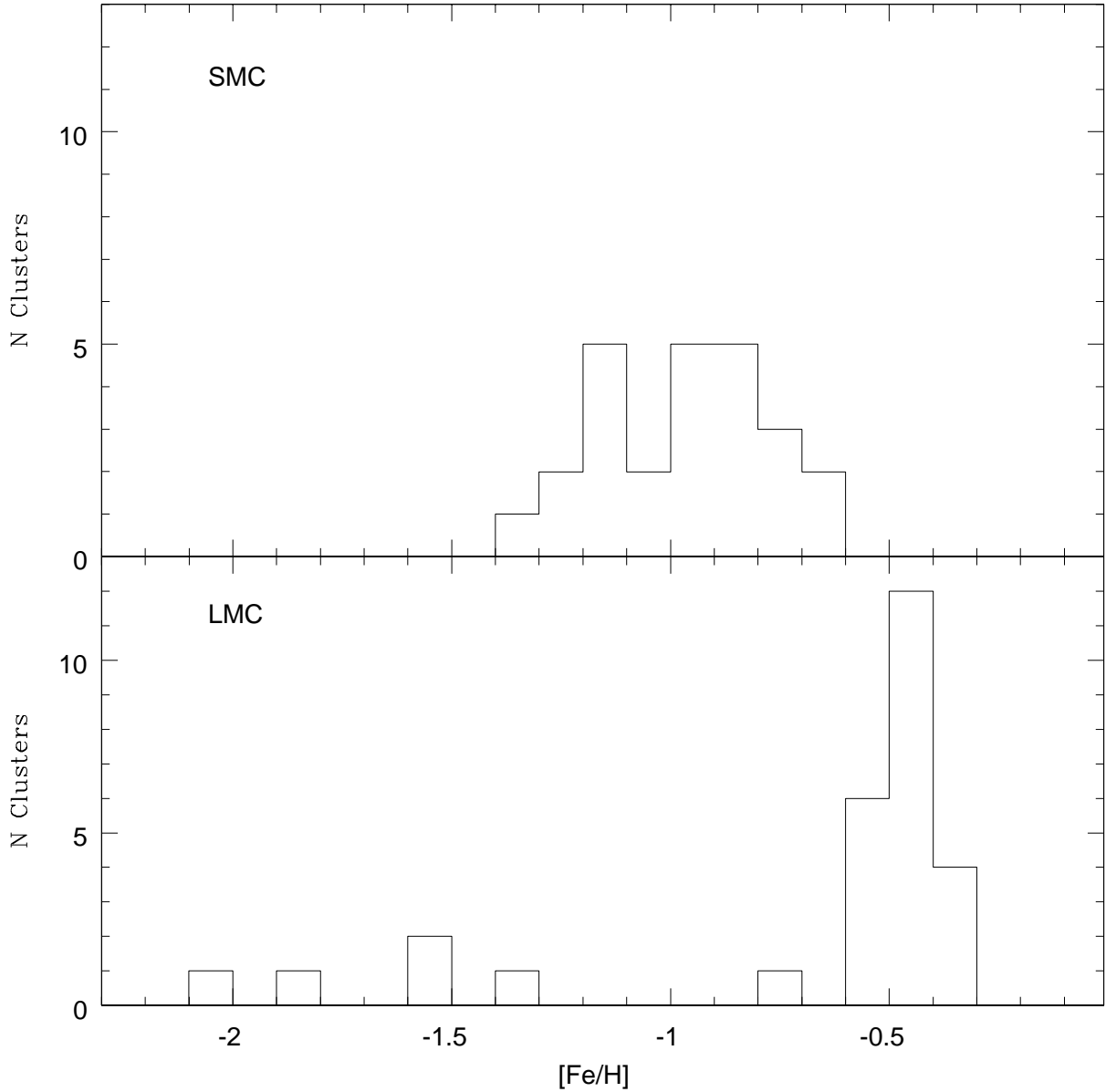}
\caption{Metallicity distribution of SMC clusters (top panel): 15 from
the present work, 6 from DH98, 3 from \citet{gl08b} and NGC\,330 \citep{gon99}.
Histogram on the bottom panel corresponds to the metallicity distribution for LMC clusters
derived from our CaT investigation (G06).}
\end{figure}

\begin{figure}
\plotone{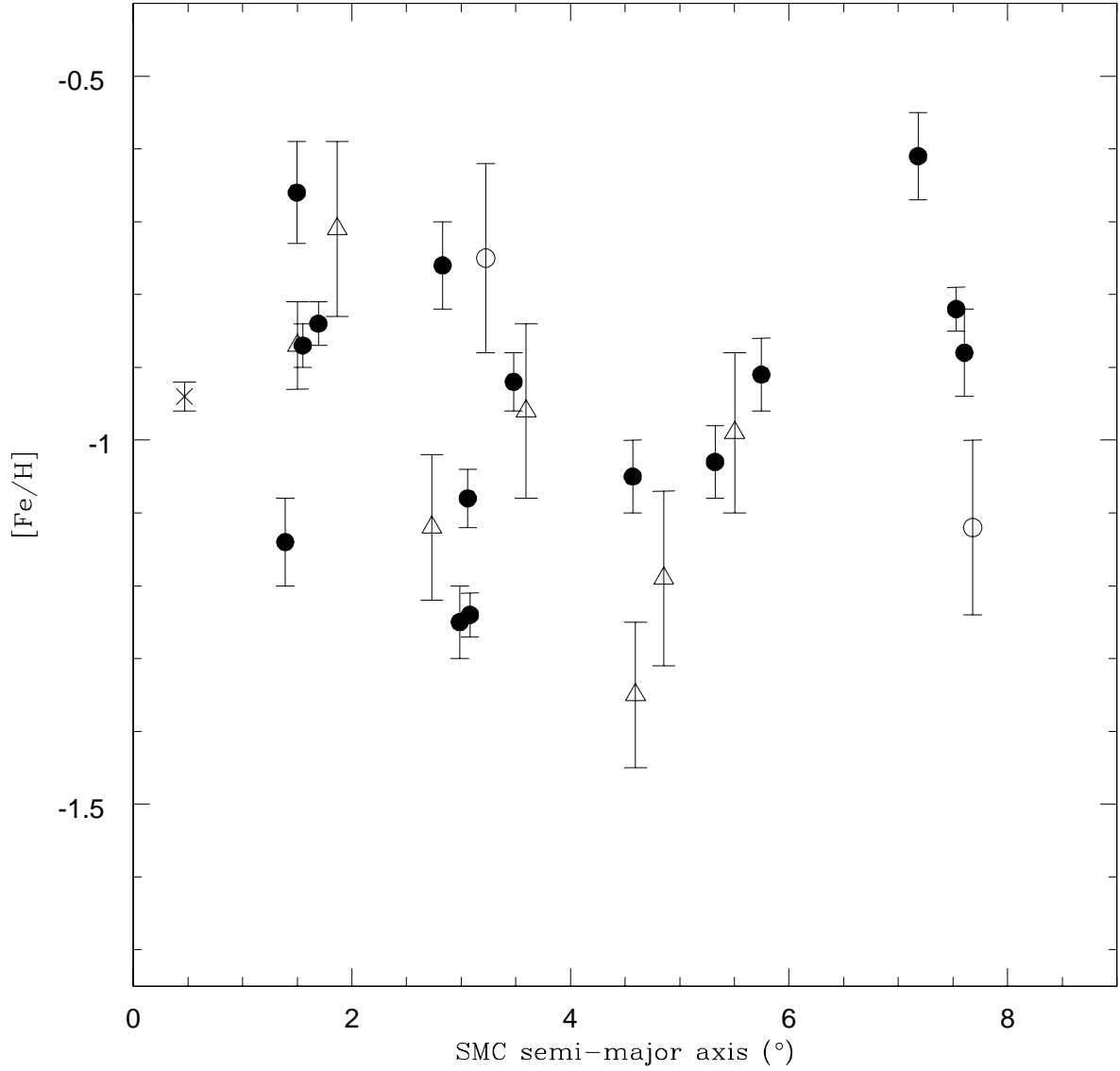}
\caption{Metallicity vs. projected radius (semi-major axis $a$) for the SMC clusters. 
Open circles represent clusters from DH98 and triangles represent clusters from \citet{gl08b}. 
Clusters from our CaT sample are represented by filled circles. NGC\,330 is shown by a cross.
No clear trend is evident.}
\end{figure}

\begin{figure}
\plotone{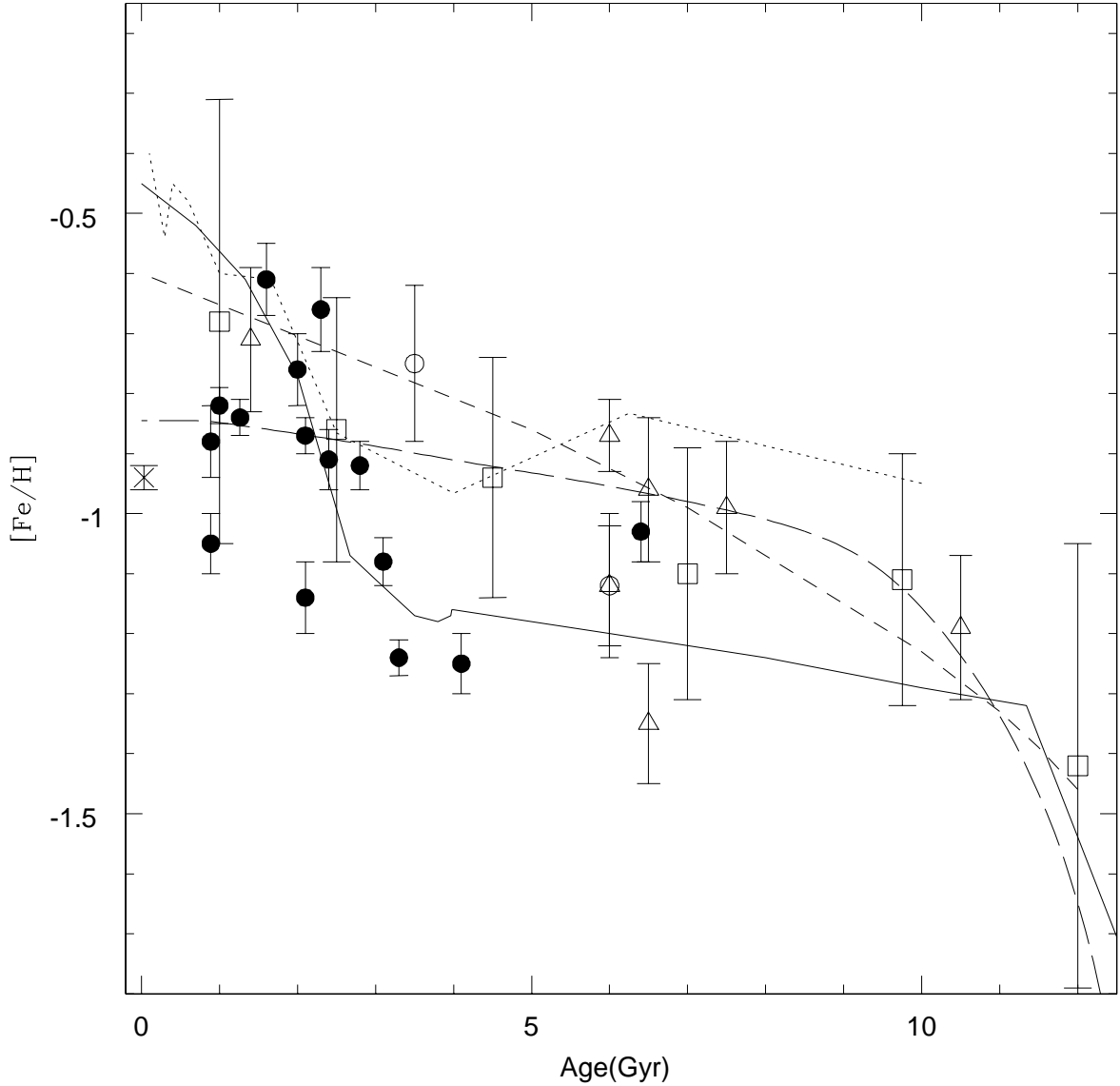}
\caption{Age-Metallicity relation. Open circles represent clusters from DH98 and triangles represent
clusters from \citet{gl08b}. Clusters of our sample are represented by filled circles. NGC\,330 is shown by a cross.
The mean metallicity in six age bins calculated by \citet{car08} are also shown (squares). 
The short dashed line represents the model of closed box continuous star formation computed by 
\citet{dch98}, the solid line corresponds to the bursting model of 
\citet{pag98} and the long dashed line shows the best-fit model derivel by \citet{car05}. 
The dotted line shows the AMR obtained by \citet{har04}. }
\end{figure}

\begin{figure}
\plotone{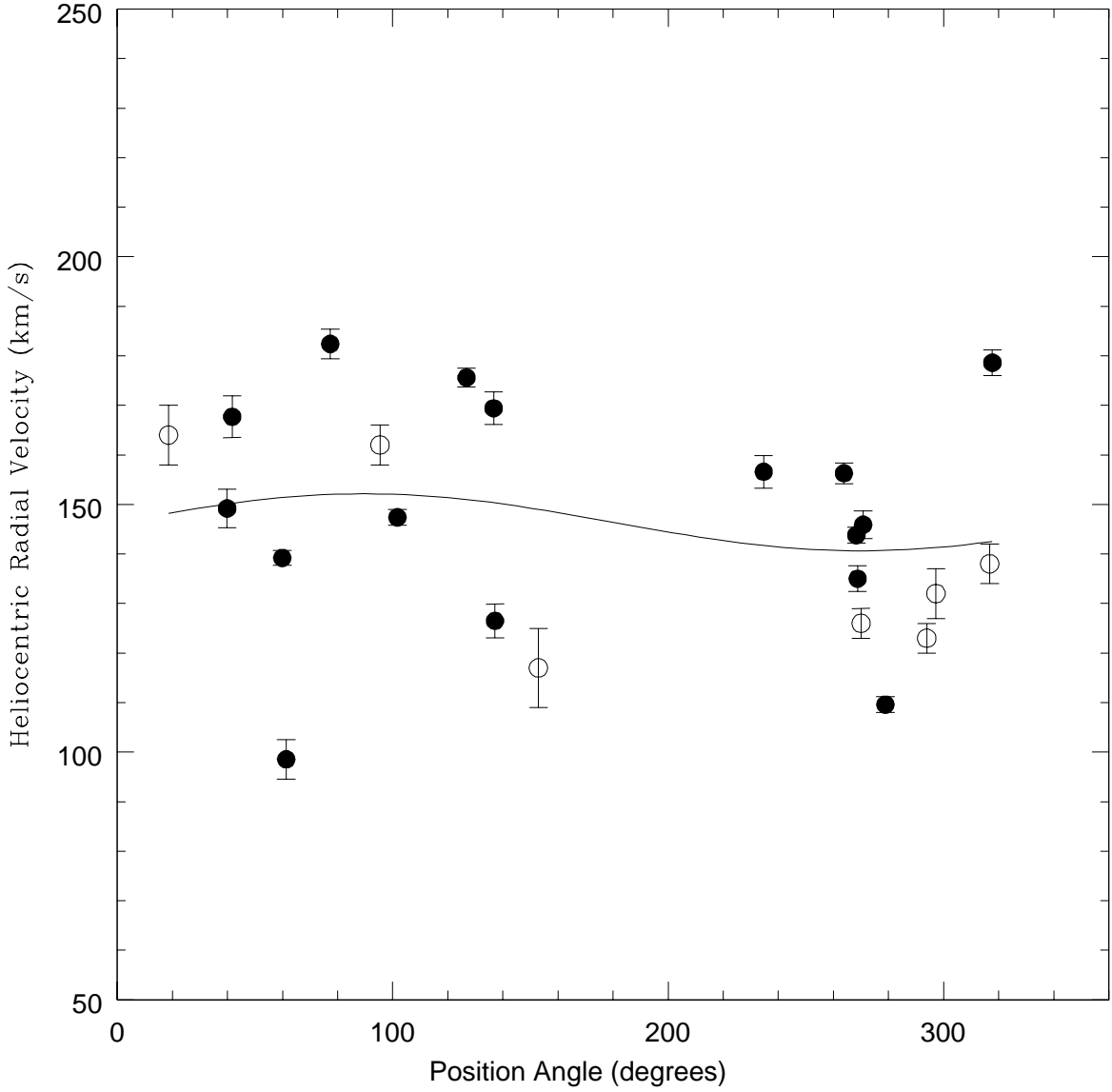}
\caption{Heliocentric radial velocity as a function of position angle on the sky. Clusters from DH98 and this
work are represented by open and filled circles, respectively. The solid line represents the best fit of a rotation curve. 
However, the curve is not
statistically better than a constant velocity.}
\end{figure}


\begin{deluxetable}{lcc}
\tablewidth{0pt}
\tablecaption{SMC Clusters}
\tablehead{
\colhead{Cluster}                & \colhead{RA (J2000.0)}  &
\colhead{Dec (J2000.0)}                                    \\
                                 & \colhead{($h$ $m$ $s$)} & 
\colhead{($^{\circ}$ $'$ $''$)}                            }
\startdata
BS\,121 = SMC\,OGLE\,237                             & 01 04 22 & -72 50 52 \\
HW\,47                                               & 01 04 04 & -74 37 09 \\
HW\,84                                               & 01 41 28 & -71 09 58 \\
HW\,86                                               & 01 42 22 & -74 10 24 \\
L\,4 = K\,1, ESO\,28-15                              & 00 21 27 & -73 44 55 \\
L\,5 = ESO\,28-16                                    & 00 22 40 & -75 04 29 \\
L\,6 = K\,4, ESO\,28-17                              & 00 23 04 & -73 40 11 \\
L\,7 = K\,5, ESO\,28-18                              & 00 24 43 & -73 45 18 \\
L\,17 = K\,13, ESO\,29-1                             & 00 35 42 & -73 35 51 \\
L\,19 = SMC\,OGLE 3                                  & 00 37 42 & -73 54 30 \\
L\,27 = K\,21, SMC\,OGLE 12                          & 00 41 24 & -72 53 27 \\
L\,72 = NGC\,376, K\,49, ESO\,29-29, SMC\,OGLE\,139  & 01 03 53 & -72 49 34 \\
L\,106 = ESO\,29-44                                  & 01 30 38 & -76 03 16 \\
L\,108                                               & 01 31 32 & -71 57 10 \\
L\,110 = ESO\,29-48                                  & 01 34 26 & -72 52 28 \\
L\,111 = NGC\,643, ESO\,29-50                        & 01 35 00 & -75 33 24 \\
\enddata
\end{deluxetable}


\begin{landscape}
\begin{deluxetable}{lccccccccc}
\tablewidth{0pt}
\tablecaption{Position and Measured Values for Member Stars}
\tablehead{
\colhead{ID}                    & \colhead{RA (J2000.0)}        &
\colhead{Dec (J2000.0)}         & \colhead{RV}                  &
\colhead{$\sigma_{RV}$}         & \colhead{v-v$_{HB}$}          &
\colhead{$\Sigma W$}             & \colhead{$\sigma_{\Sigma W}$}&
\colhead{[Fe/H]}                 & \colhead{$\sigma_{[Fe/H]}$} \\
                                & \colhead{($h$ $m$ $s$)}       &
\colhead{($^{\circ}$ $'$ $''$)} & \colhead{(km s$^{-1}$)}       &
\colhead{(km s$^{-1}$)}         & \colhead{(mag)}               &
\colhead{(\AA)}                 & \colhead{(\AA)}               &
                                &                               }
\startdata
BS\,121-2  & 01 04 17.53 & -72 50 20.00 & 173.7 & 5.3 & -0.66 & 6.58 & 0.27 & -0.758 & 0.141 \\
BS\,121-4  & 01 04 21.15 & -72 50 33.60 & 175.0 & 5.3 & -1.00 & 7.46 & 0.17 & -0.529 & 0.128 \\
BS\,121-5  & 01 04 25.85 & -72 50 21.82 & 157.8 & 5.4 & -1.00 & 6.42 & 0.21 & -0.907 & 0.124 \\
BS\,121-7  & 01 04 24.05 & -72 51 04.80 & 175.1 & 5.3 & -1.03 & 7.39 & 0.15 & -0.562 & 0.123 \\
BS\,121-12 & 01 04 32.46 & -72 51 35.92 & 156.9 & 5.3 & -0.61 & 7.16 & 0.20 & -0.534 & 0.133 \\
\enddata
\tablecomments{Table 2 is published in its entirety in the
electronic edition of the {\it Astronomical Journal}.  A portion is
shown here for guidance regarding its form and content.}

\end{deluxetable}
\end{landscape}


\begin{deluxetable}{lcccccrc}
\tablewidth{0pt}
\tablecaption{Derived SMC Cluster Properties}
\tablehead{
\colhead{Cluster}       & \colhead{n}                 &
\colhead{RV}            & \colhead{$\sigma_{RV}$}     &
\colhead{[Fe/H]}        & \colhead{$\sigma_{[Fe/H]}$} &
\colhead{P.A.}          & \colhead{$a$}               \\
                        &                             &
\colhead{(km s$^{-1}$)} & \colhead{(km s$^{-1}$)}     &
\colhead{(dex)}         & \colhead{(dex)}             &
\colhead{($^{\circ}$)}  & \colhead{($^{\circ}$)}      }
\startdata
BS\,121 & 5 & 167.7 & 4.2 & -0.66 & 0.07 & 41.73  & 1.496 \\
HW\,47  & 4 & 126.5 & 3.4 & -0.92 & 0.04 & 137.14 & 3.502 \\
HW\,84  & 4 & 139.2 & 1.5 & -0.91 & 0.05 & 59.90  & 5.513 \\
HW\,86  & 4 & 147.4 & 1.6 & -0.61 & 0.06 & 101.70 & 7.345 \\
L\,4    & 9 & 143.8 & 1.6 & -1.08 & 0.04 & 268.34 & 3.265 \\
L\,5    & 5 & 156.6 & 3.3 & -1.25 & 0.05 & 234.70 & 3.092 \\
L\,6    & 7 & 145.9 & 2.8 & -1.24 & 0.03 & 270.81 & 3.124 \\
L\,7    & 7 & 135.0 & 2.6 & -0.76 & 0.06 & 268.79 & 2.888 \\
L\,17   & 8 & 109.6 & 1.6 & -0.84 & 0.03 & 278.90 & 1.718 \\
L\,19   & 7 & 156.3 & 2.1 & -0.87 & 0.03 & 263.83 & 1.564 \\
L\,27   & 7 & 178.6 & 2.6 & -1.14 & 0.06 & 317.73 & 1.392 \\
L\,72   & 5 & 149.2 & 3.9 & \nodata & \nodata & 39.88  & 1.410 \\
L\,106  & 7 & 169.4 & 3.3 & -0.88 & 0.06 & 136.64 & 7.877 \\
L\,108  & 6 & 98.55 & 4.0 & -1.05 & 0.05 & 61.37  & 4.460 \\
L\,110  & 9 & 182.4 & 3.0 & -1.03 & 0.05 & 77.32  & 5.323 \\
L\,111  & 8 & 175.6 & 1.9 & -0.82 & 0.03 & 126.81 & 7.830 \\
\enddata
\end{deluxetable}


\begin{deluxetable}{lccc}
\tablewidth{0pt}
\tablecaption{Photometric Cluster Metallicities and Ages}
\tablehead{
\colhead{Cluster} & \colhead{Metallicity} &
\colhead{Age}     & \colhead{Reference}   \\
                  & \colhead{(dex)}       & 
\colhead{(Gyr)}   & \colhead{Met., Age}   }
\startdata
BS\,121 & -1.2  $\pm$ 0.4              & 2.3                      & 1, 1      \\
HW\,47  & -1.0  $\pm$ 0.4              & 2.8                      & 1, 1      \\
HW\,84  & -1.2  $\pm$ 0.4              & 2.4                      & 1, 1      \\
HW\,86  & -0.75 $\pm$ 0.4              & 1.6                      & 1, 1      \\
L\,4    & -0.9  $\pm$ 0.2              & 3.1                      & 1, 1      \\
L\,5    & -1.2  $\pm$ 0.2              & 4.1                      & 1, 1      \\
L\,6    & -0.9  $\pm$ 0.2              & 3.3                      & 1, 1      \\
L\,7    & -0.6  $\pm$ 0.2              & 2.0                      & 1, 1      \\
        & Z/Z$_\sun$ = -1.1 $\pm$ 0.2  & \nodata                 & 2,\nodata \\
L\,17   & \nodata                      & 1.26                     & \nodata, 4\\
L\,19   & -0.75 $\pm$ 0.2              & 2.1                      & 1,1       \\
L\,27   & -1.3 $\pm$ 0.3               & 2.1                      & 1,1       \\
L\,72   & \nodata                      & 0.025 $\pm$ 0.010        & \nodata, 5\\
L\,106  & \nodata                      & 0.89 $^{+0.23}_{-0.10}$  & \nodata, 6\\
L\,108  & \nodata                      & 0.89 $^{+0.37}_{-0.18}$  & \nodata, 6\\
L\,110  & -1.15 $\pm$ 0.25             & 6.4 $\pm$ 1.1            & 3, 3      \\
L\,111  & Z/Z$_\sun$ = -0.6 $\pm$ 0.25 & 1.00 $^{+0.26}_{-0.21}$  & 2, 6      \\
\enddata
\tablerefs{
(1) \citealt{pi05a}; (2) \citealt{bic86}; 
(3) \citealt{pi07b}; (4) \citealt{raf05};
(5) \citealt{pi07a}; (6) \citealt{pi07c}.}
\end{deluxetable}


\begin{deluxetable}{lccc}
\tablewidth{0pt}
\tablecaption{Aditional Cluster Samples}
\tablehead{
\colhead{Cluster} & \colhead{Metallicity} &
\colhead{Age}     & \colhead{Reference} \\
                  & \colhead{(dex)}       &
\colhead{(Gyr)}   & \colhead{Met., Age} }
\startdata
L\,11  & -0.75 $\pm$ 0.13  & 3.5 $\pm$ 1.0 & 1, 1 \\ 
L\,113 & -1.12 $\pm$ 0.12  & 6.0 $\pm$ 1.0   & 1, 1 \\
NGC\,121 & -1.19 $\pm$ 0.12 & 10.5 $\pm$ 0.5 & 1, 2 \\
L\,1  & -0.99 $\pm$ 0.11 & 7.5 $\pm$ 0.5 & 1, 2 \\
K\,3 & -0.96 $\pm$ 0.12  & 6.5 $\pm$ 0.5 & 1, 2 \\
NGC\,339 & -1.12 $\pm$ 0.10 & 6.0 $\pm$ 0.5 & 1, 2 \\
NGC\,416 & -1.00 $\pm$ 0.13 & 6.0 $\pm$ 0.8 & 2, 2 \\
L\,38 & -1.59 $\pm$ 0.10 & 6.5 $\pm$ 0.5 & 2, 2 \\
NGC\,419 & -0.67 $\pm$ 0.12 & 1.4 & 2, 2 \\
NGC\,330 & -0.94 $\pm$ 0.02 & 0.03 & 3, 4 \\

\enddata
\tablerefs{
(1) \citealt{dch98}; (2) \citealt{gl08b};
(3) \citealt{gon99}; (4) WEBDA
}
\end{deluxetable}

\end{document}